\documentclass[twocolumn]{aa}
\usepackage[varg]{txfonts}
\usepackage{color}
\usepackage{amsmath}
\usepackage{mathtools}
\usepackage{float}
\usepackage{natbib}
\usepackage{graphicx}
\usepackage{graphicx}
\usepackage{ulem}
\usepackage{lscape}
\usepackage{xspace}
\usepackage{changes}

\makeatletter
\renewcommand*\aa@pageof{, page \thepage{} of \pageref*{LastPage}}
\makeatother
\newcommand\blfootnote[1]{%
  \begingroup
  \renewcommand\thefootnote{}\footnote{#1}%
  \addtocounter{footnote}{-1}%
  \endgroup
}

\usepackage[colorlinks = true,
            linkcolor = blue,
            urlcolor  = blue,
            citecolor = blue,
            anchorcolor = blue]{hyperref}
\definecolor{magenta}{rgb}{1.0, 0.0, 1.0}
\definecolor{mygray}{gray}{0.6}
\definecolor{calpolypomonagreen}{rgb}{0.12, 0.3, 0.17}
\definecolor{darkred}{rgb}{0.55, 0.0, 0.0}

\def\h2o{H$_2$O}

\def\kgm3{kg\,m$^{-3}$}
\def\Jkg3{J\,kg$^{-3}$}
\begin{document} 

%%%%%%%%%%%%%%%%%%%%%%%%%       Title page      %%%%%%%%%%%%%%%%%%%%%%%%%
    \title{Prograde spin-up during gravitational collapse}
   
   \author{Rico G. Visser \inst{1 \star}
          \and
           Marc G. Brouwers \inst{2 \star}
          }

   \institute{Anton Pannekoek Institute, University of Amsterdam, Science Park 904, Amsterdam, The Netherlands
   \and 
   Institute of Astronomy, University of Cambridge, Madingley Road, Cambridge CB3 0HA, United Kingdom
   \\
              \email{r.g.visser@uva.nl,mgb52@cam.ac.uk}
             }

  \abstract{
Asteroids, planets, stars in some open clusters, as well as molecular clouds appear to possess a preferential spin-orbit alignment, pointing to shared processes that tie their rotation at birth to larger parent structures. We present a new mechanism that describes how collections of particles or 'clouds' gain a prograde rotational component when they collapse or contract while subject to an external, central force. The effect is geometric in origin, as relative shear on curved orbits moves their shared center-of-mass slightly inward and toward the external potential during a collapse, exchanging orbital angular momentum into aligned (prograde) rotation. We perform illustrative analytical and N-body calculations to show that this process of prograde spin-up proceeds quadratically in time ($\delta L_\mathrm{rot} \propto t^2$) until the collapse nears completion. The total rotational gain increases with the size of the cloud prior to its collapse: $\delta L_\mathrm{rot}/L_\mathrm{H} \propto (R_\mathrm{cl}/R_\mathrm{H})^5$, and typically with distance to the source of the potential ($L_\mathrm{H}\propto r_0)$. For clouds that form at the interface of shear and self-gravity ($R_\mathrm{cl} \sim R_\mathrm{H}$), prograde spin-up means that even setups with large initial retrograde rotation collapse to form prograde-spinning objects. Being a geometric effect, prograde spin-up persists around any central potential that triggers shear, even those where the shear is strongly retrograde. We highlight an application to the Solar System, where prograde spin-up can explain the frequency of binary objects in the Kuiper belt with prograde rotation.
}
   
   \maketitle

%

%

%%%%%%
\section{Introduction}

\begin{figure*}[t!] 
\centering
\includegraphics[width=1\textwidth]{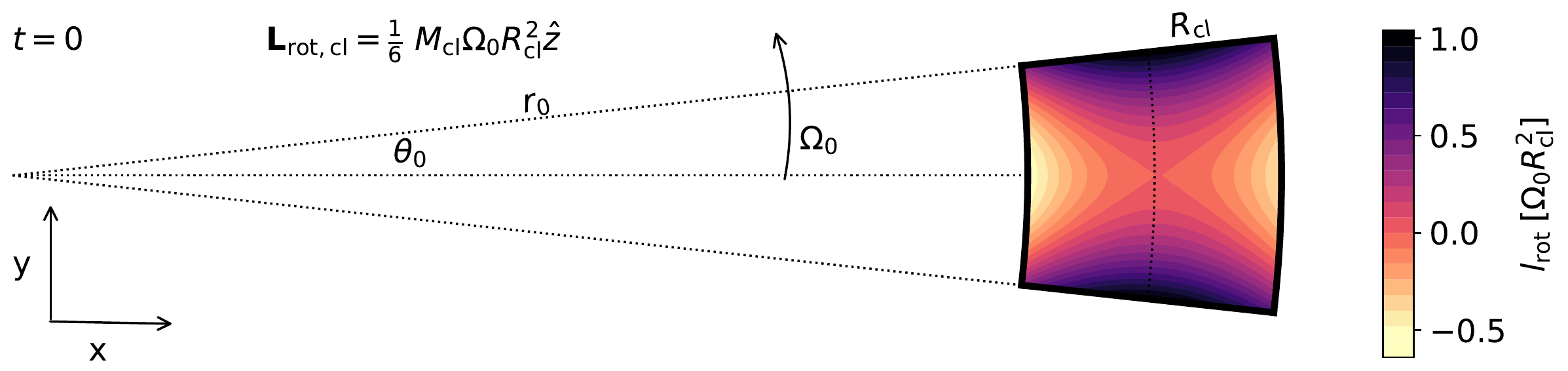} \label{Fig:analytical_schematic}
\caption{Schematic setup of our analytical calculation (Sect. \ref{sect:analytical}). We consider a two-dimensional cloud of mass $M_\mathrm{cl}$ in a stationary frame, whose origin lies on the central mass $M_\mathrm{C} \gg M_\mathrm{cl}$ (where the dotted lines cross). The initial shape of the cloud is that of a partial annulus, bounded by inner and outer circles with radii $r_0 \pm R_\mathrm{cl}$ and radial lines at angles $\pm \theta_0$. We follow the position ($\bar{x}_\mathrm{cl}, \bar{y}_\mathrm{cl}$) and velocity ($\bar{v}_\mathrm{x,cl}, \bar{v}_\mathrm{y,cl}$) of the cloud's center-of-mass over time and calculate the specific rotational angular momentum of cloud particles ($\hat{z}$) relative to this point as $l_\mathrm{rot} = (x v_\mathrm{y} - y v_\mathrm{x}) - (\bar{x}_\mathrm{cl} \bar{v}_\mathrm{y,cl} - \bar{y}_\mathrm{cl} \bar{v}_\mathrm{x,cl})$, shown in the color plot at $t=0$. The initial velocities of the cloud particles correspond to circular, Keplerian motion. Particles that start on the x-axis initially contribute retrograde rotation due to shear, whereas particles on the y-axis contribute prograde rotation due to curvature. The integrated result is a net initial prograde rotational angular momentum that increases over time if the cloud orbits freely (See Eq. \ref{eq:L_rot_analytical}).}
\label{fig:arc}
\end{figure*}

Across a wide range of astrophysical scales, from asteroids to stars and beyond, objects of many kinds tend to form via the gravitational collapse of larger structures. The rotation of these objects, which tends to manifest as increased velocities when their moments of inertia decrease, is a key tool to investigate the physics of their formation. When inferring rotational trends from observations, it is often difficult to disentangle the system's rotation at birth from its later dynamical evolution. In some cases, however, systems with largely primordial rotational distributions still contain important historical clues about the processes by which they formed. \blfootnote{$\star$ Both authors contributed equally to this work. Author ordering was determined randomly.}

At the smallest scale of relevance here, asteroids and comets are long since thought to form in gravitational collapses within proto-planetary disks \citep[e.g.][]{Goldreich1973, Youdin2002, Morbidelli2009}, via a small-scale mechanism now known as the streaming instability \citep{youdingoodman2005, JohansenEtal2007, JohansenEtal2009, Simon2016, Schafer2017}. Whether traces of their primordial spins remain, depends chiefly on the class and size of asteroids that are studied. At the lower end of the size distribution, smaller asteroids have a largely isotropic spin distribution, with traces of their prior rotation wiped out by intense collisional processing \citep[e.g.][]{Davis1989, Bottkeetal2005a, Bottke2005b, Pan2005, Fraser2009} or by the YORP effect \citep{Pravecetal2008, Medeirosetal2018, Rubincam2000}. Similarly, the largest bodies like Ceres and Vesta may have accreted most of their prograde spin during later stages by pebble accretion \citep{Johansen2010, Visseretal2020}. In between, however, an intermediate class of asteroids with diameters between $\sim 100-500$ km likely still contains key information about their rotation at birth \citep{Steinberg2015}. Within the Solar System, the most dynamically pristine sub-set of these objects belong to the group of 'Cold Classicals' that reside in the Kuiper belt. Besides being characterized by low inclinations and eccentricities, this group contains a high fraction of binary pairs \citep{Noll2008, Fraser2017, Grundy2019} that often have strong color correlations \citep{Benecchi2009, Benecchi2011, Marsset2020}, reinforcing the idea that they formed in a single gravitational collapse - rather than by later capture. Interestingly, many of the binaries are found to have mass ratio's near unity - which means that the collections of particles or 'clouds' from which they formed, must have contained substantial amounts of rotational angular momentum \citep{Nesvorny2010, Robinson2020}. Their binary orbits are not distributed isotropically and express a preferential (prograde) alignment with their center-of-mass orbit around the Sun \citep{Grundy2019}. Recently, hydrodynamic simulations have shown that such a distribution can indeed arise as the result of streaming instabilities \citep{Nesvorny2019, Nesvorny2021}, but the underlying physical origin of the strong spin-orbit alignment remains poorly understood.

The trend of spin-orbit alignment continues at the scale above asteroids, where all the planets in the Solar System except Venus and Uranus rotate in the prograde direction. Theoretically, the prograde rotation of gas giants is readily explained in the context of core accretion by the torque during runaway gas accretion \citep{Machida2008, Dittmann2021} but systematic explanations of terrestrial spin have proven more problematic \citep{LissauerSafronov1991}. Planetesimal accretion is found to deliver insufficient rotation to the accreting planet \citep{LissauerKari1991, DonesTremaine1993}, whereas later giant impacts fail to produce an anisotropic distribution \citep{Safronov1966, DonesTremaine1993i}. Pebble accretion could potentially provide a large, systematically prograde spin to planets \citep{Johansen2010, Visseretal2020}, although this remains to be validated for accretion onto proto-planets with substantial envelopes, where pebbles face intense drag and sublimation \citep{Alibert2017, Brouwers2021,Johansen2020}.

In the alternative theories to core accretion, planetary rotation again finds its origin in gravitational collapse. Terrestrial cores have been proposed to form out of dense pebble traps at the edges of dead zones in the inside-out mechanism \citep{Chatterjee2014, Hu2016, Hu2018, Mohanty2018, Cai2022}, whereas gas giants can form directly from gravitational instabilities of cold gas \citep[e.g.][]{Kuiper1951, Cameron1978, Boss1998, Boss2021}. The obliquities of these larger planets on wide orbits are just beginning to be measured. Observations of 2M0122b seem to tentatively indicate a large obliquity \citep{Bryan2020, Bryan2021}, while HD 106906 b more certainly spins on its side \citep{Bryan2021}. Recent SPH and hydrodynamic simulations show that gravitational cloud collapse in gravito-turbulent discs can yield obliquities up to 90 degrees \citep{Jennings2021}, although most objects seem to form with more alignment \citep{Hall2017}. For now, more observations beyond the two known cases are required to resolve the statistical spin distribution of this class of planets.

Most stars form in clusters by the collapse of dense sub-structures within molecular clouds, that often fragment into several thousands of stars per cluster \citep[e.g.][]{McKee2007, Lee2012, Lee2016b}. The mutual alignment of stars in these clusters has been studied in a limited context, with differing results. In the first spectroscopic studies of the young, low-mass open clusters Pleiades and Alpha Per, no spin alignment was found in the population of G,K-, or M-type stars \citep{Jackson2010, Jackson2018}. This seems consistent with the older idea that stellar spins depend intimately on the local turbulence \citep[e.g.][]{Fleck1981, Belloche2013}, which itself is ultimately induced by differential galactic rotation \citep{Renaud2013, Rey2015, Reyetal2018}. Recently, however, astro-seismological measurements of red giants in the higher-mass clusters NGC 6791 and NGC 6819 were found to exhibit a strong inter-cluster spin alignment \citep{Corsaroetal2017}. High-resolution simulations indicate that such spin-alignment can arise for stars above 0.7 $\mathrm{M_\odot}$ when their natal star-forming clump contains more than 50\% of its kinetic energy in rotation \citep{Lee2016a, Corsaroetal2017}. Although this rotational support might be rare \citep{Caselli2002,Pirogov2003}, mutual stellar spin alignment in such cases points to a spin-orbit alignment within the cluster, similar to the prograde rotation of asteroids and planets around a central star. Finally, we add that in the well-studied galaxies M33 and M51, the larger molecular clouds that host the star-forming clumps also seem to possess structural prograde rotation around the galaxy itself \citep{Braine2018, Braine2020}.

The preceding overview reveals a universal trend; objects that form via a gravitational collapse in the presence of an external gravitational field tend to exhibit a preferential alignment between their spins and orbital motions. Remarkably, the fact that such a spin-orbit alignment exists across different scales has received little scientific attention and no satisfactory common explanation has yet been suggested. Perhaps the most relevant contribution to the topic is that of \citet{Mestel1966}, who considered the instantaneous condensation of patches from a differentially rotating disk ($\Omega(r) \propto r^n$). His work showed that any closed, two-dimensional patch of a disk with an increasing rotation curve ($n\geq -1$) always contains prograde spin about its center-of-mass, whereas patches from a Keplerian disk ($n=-3/2$) can yield spin in either direction, depending on their shape.
However, in many cases, the collapse will not proceed from a fully ordered state. If the relevant clouds are formed by a process in which turbulent motions play an important role, the initial rotational setup of the collapsing cloud will be much more randomized, making it unclear from the potential alone what the spin direction will be. In addition, \citet{Shmidt1957} realized that the center-of-mass of a patch - and therefore its orbital angular momentum - could evolve during its collapse and that the final rotational direction would formally be set by the conservation of total angular momentum and energy. In his work, prograde rotation was interpreted as the result of a collapse with sufficient thermal energy loss. These calculations were later extended by \citet{Safronov1962, Safronov1972} but ultimately proved inconclusive due to the problem of unknown thermal losses.

In this work, we provide a different, geometrically motivated approach to the rotational evolution of collapsing clouds in orbit around an external potential, most similar to the tidal torque theory that seeks to explain galactic spins \citep{Hoyle1951, Peebles1969}. We show that because particles that orbit in any non-rigid cloud shear away from one another over time - and do so on curved paths - their combined center-of-mass moves toward the source of the external potential during a gravitational collapse. The orbital angular momentum that is thus liberated, adds a prograde component to the spin of the object that forms. This mechanism of prograde spin-up is most effective when the clouds are low in density prior to their collapse and when individual particles are on circular orbits, although the effect persists across a wide range of setups. Due to its universal applicability, we suggest that prograde spin-up during a gravitational collapse contributes to the ubiquity of spin-orbit alignment on different scales.

The structure of the paper is organized as follows. We begin in Sect. \ref{sect:analytical} with an illustrative analytical calculation of prograde spin-up with two-dimensional, circular orbits around a point source. In Sect. \ref{sect:NumCC}, we numerically verify the main analytical trends by including self-gravity and present a more visual analysis by showing the spin-up of a cloud without any initial rotation. We apply the mechanism of prograde spin-up to the formation of binary asteroids/comets via streaming instability in Sect. \ref{sect:SI}. Finally, we discuss the broader implications of our findings in Sect. \ref{sect:Discussion} and conclude in Sect. \ref{sect:Conclusions}.

%%%%%%
\section{Analytical evaluation of prograde spin-up}\label{sect:analytical}
In this section, we perform a calculation without self-gravity to illustrate how shear can transfer orbital angular momentum to rotation based on the orbital geometry. The idea behind this calculation is that objects form at the centers-of-mass of collapsing clouds, which orbit in the potential of larger objects or structures. While the non-linear nature of a cloud collapse in the presence of an external potential is generally unsuitable for analytical modeling, the main trends can be elucidated by following cloud particles without treating self-gravity, at least while the collapse is not yet in full swing. The setup of this analytical calculation is sketched in Fig. \ref{Fig:analytical_schematic}. We consider a two-dimensional cloud with a uniform density $\sigma_\mathrm{cl}$ that integrates to a total mass $M_\mathrm{cl}$. The particles in the cloud move around a stationary central mass $M_\mathrm{C} \gg M_\mathrm{cl}$ on circular orbits with angular velocity $\Omega = \sqrt{G M_\mathrm{C} / r^3}$ at distance $r$. We conveniently let the initial shape of the cloud be that of a partial annulus, bounded by inner and outer circles at $r_0 \pm R_\mathrm{cl}$ and angles $\pm \theta_0$, with $R_\mathrm{cl} = \theta_0 r_0$, such that the cloud consists of radially separated differential arcs with length $l_\mathrm{arc}(r) = 2R_\mathrm{cl}r/r_0$.

\subsection{Position and velocity of shearing radial arcs}
First, we calculate the positions and velocities of the different radial arcs. Their angular velocities vary slightly depending on their radial separation, causing deviations between arcs that grow over time. In a stationary frame, the upper ($\theta_{+}$) and lower ($\theta_{-}$) bounding angles of an arc are given by:
\begin{equation}
    \theta_\mathrm{\pm} = \pm \theta_0 + \Omega_0 \left(\frac{r}{r_0}\right)^{-\frac{3}{2}} t,
\end{equation}
where $t$ is time after initiation. The center-of-mass of an arc can be calculated in Cartesian coordinates with their origin on $M_\mathrm{C}$ ($\bar{x}_\mathrm{arc}, \bar{y}_\mathrm{arc}$) to lie at:
\begin{subequations}
\begin{align}
    \bar{x}_\mathrm{arc} &= \frac{1}{2\theta_0} \int_{\theta_{-}}^{\theta_{+}} r \; \mathrm{cos} \left(\theta '\right)d\theta ', \\
    &= \frac{r \; \mathrm{sin}\left(\theta_0\right) \; \mathrm{cos}\left[\Omega_0\left(\frac{r}{r_0}\right)^{-\frac{3}{2}} t\right]}{\theta_0}, \label{eq:x_arcs}
\end{align}
\end{subequations}
\begin{subequations}
\begin{align}
    \bar{y}_\mathrm{arc} &= \frac{1}{2\theta_0} \int_{\theta_{-}}^{\theta_{+}} r \; \mathrm{sin \left(\theta '\right)d\theta '} \\
    &= \frac{r \; \mathrm{sin}\left(\theta_0\right) \; \mathrm{sin}\left[\Omega_0\left(\frac{r}{r_0}\right)^{-\frac{3}{2}} t\right]}{\theta_0}. \label{eq:y_arcs}
\end{align}
\end{subequations}
As expected, it is initially located on the x-axis and subsequently orbits in a circular motion as a function of time. Trivially, its velocity components follow from the time derivatives as:
\begin{equation}\label{eq:v_arcs}
    \bar{v}_\mathrm{x, arc} = -\Omega_0 \left(\frac{r}{r_0}\right)^{-\frac{3}{2}} \bar{y}_\mathrm{arc} \;\;,\;\; \bar{v}_\mathrm{y, arc} = \Omega_0 \left(\frac{r}{r_0}\right)^{-\frac{3}{2}} \bar{x}_\mathrm{arc}.
\end{equation}

\subsection{Trajectory of the cloud's center-of-mass}
\begin{figure}[t!] 
\centering
\includegraphics[width=.49\textwidth]{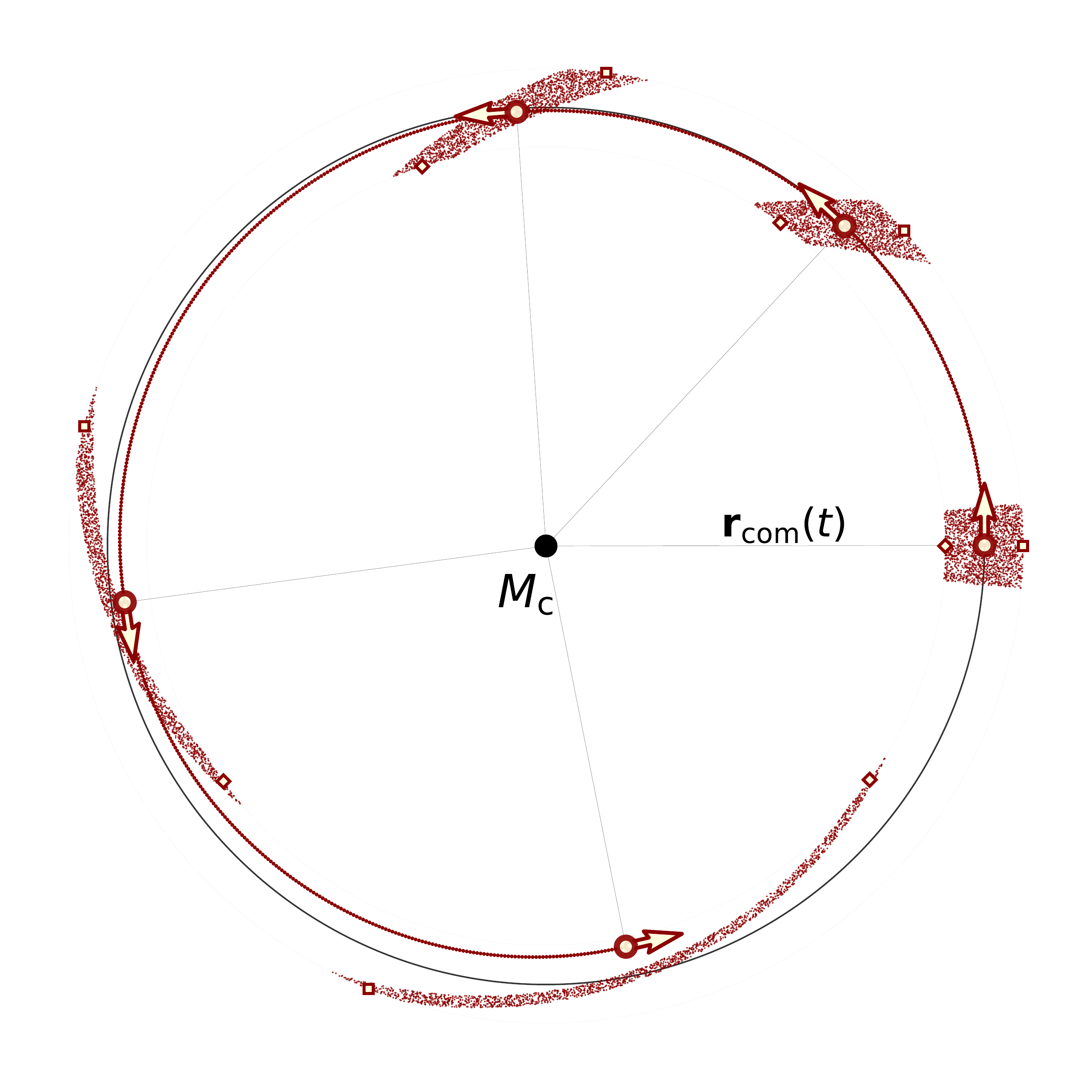}
\caption{Illustration of the inward center-of-mass shift in our analytical example. The particles follow circular orbits around a central mass without mutual interactions. As the particles shear out over time, the ensemble deforms into an increasingly extended single arc, shifting the center-of-mass position (open circular dot) closer to the orbital midpoint.}
\label{fig:cloudsketch}
\end{figure}
The calculation becomes interesting when we combine the positions of different arcs to find the center-of-mass of the whole cloud. We first note that without self-gravity, its area, being the integral over the differential arcs, remains constant over time and follows from subtracting the two partial disk areas as $A_\mathrm{cl} = 4 R_\mathrm{cl}r_0\theta_0 = 4 R_\mathrm{cl}^2$. The position ($\bar{x}_\mathrm{cl}, \bar{y}_\mathrm{cl}$) and velocity ($\bar{v}_\mathrm{x,cl}, \bar{v}_\mathrm{y,cl}$) of the cloud's center-of-mass can be found by radially integrating over Eqs. \ref{eq:x_arcs}, \ref{eq:y_arcs}, and \ref{eq:v_arcs}, which yields:
\begin{subequations}
\begin{align}
    \bar{x}_\mathrm{cl}(t) &= \frac{1}{A_\mathrm{cl}} \int_{r_0-R_\mathrm{cl}}^{r_0+R_\mathrm{cl}}\bar{x}_\mathrm{arc}(r) l_\mathrm{arc}(r) dr \\
    &\simeq r_0 \; \mathrm{cos}\left(\Omega_0 t\right) \nonumber \\ 
    &+ \frac{R_\mathrm{cl}^2}{24 r_0} \left[\left(4 - 9\Omega_0^2 t^2 \right) \mathrm{cos}\left(\Omega_0 t\right) + 9 \Omega_0 t \; \mathrm{sin}\left(\Omega_0 t\right)\right], \label{eq:x_cloud}
\end{align}
\end{subequations}
\begin{figure*} 
\includegraphics[width=.49\textwidth]{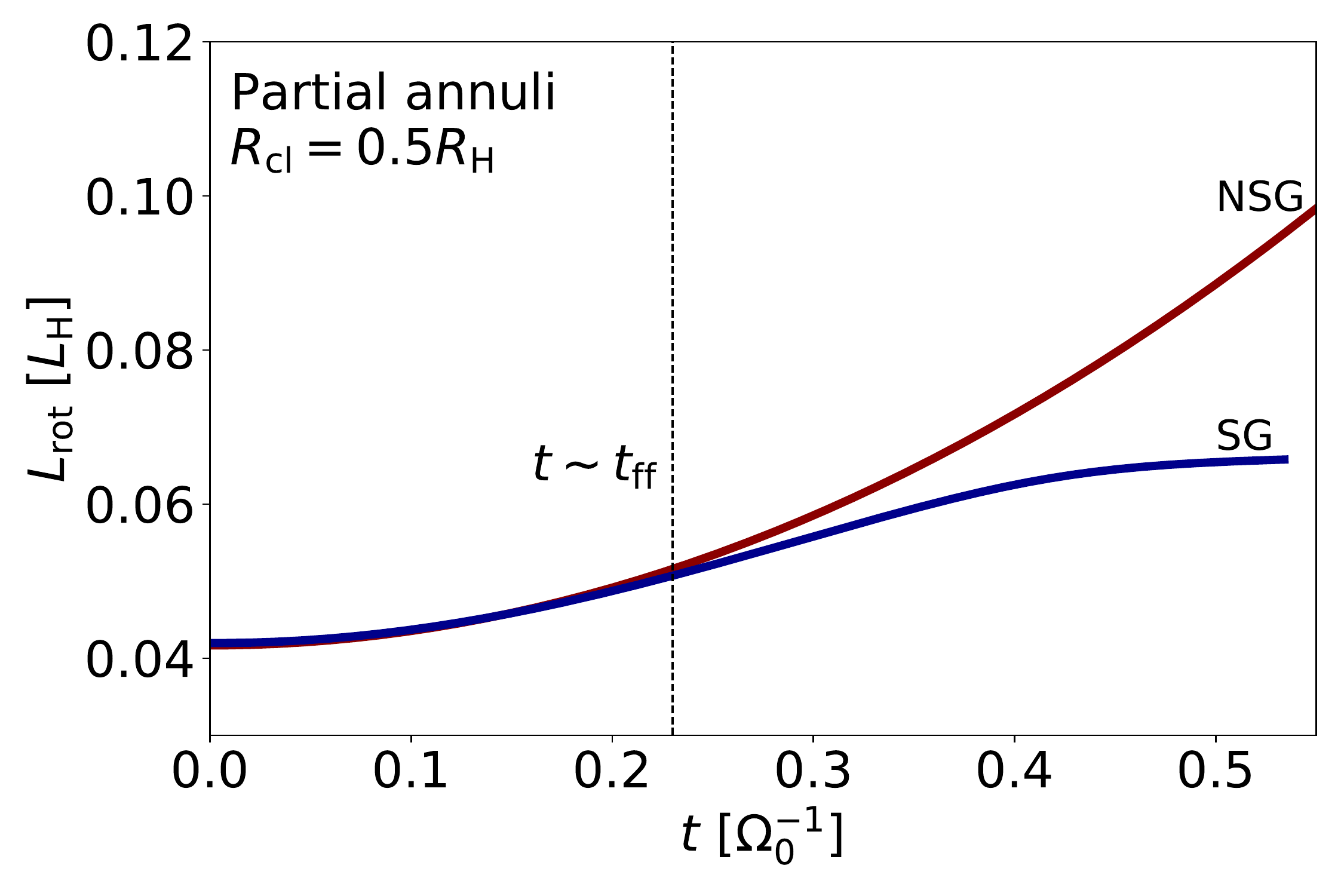}\
\includegraphics[width=.49\textwidth]{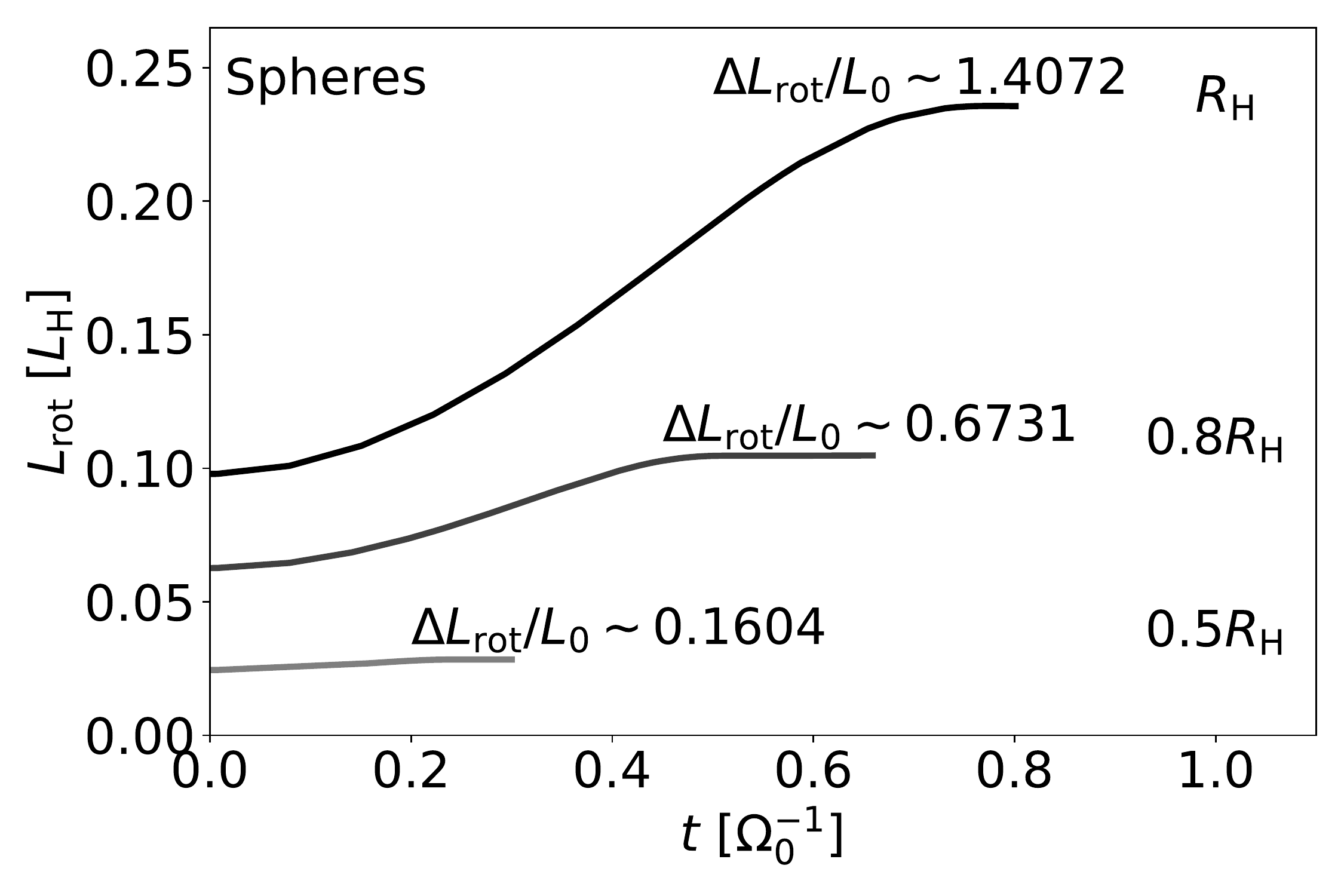}
\caption{Left: Simulated prograde spin-up of clouds in the shape of partial annuli, that orbit around a central mass with (blue, SG) and without (red, NSG, Eq. \ref{eq:dL}) self-gravity enabled. When self-gravity and collisions are included, the prograde spin-up levels off around the free-fall timescale (plotted here for an equivalent sphere), when cloud shear comes to a halt. Right: The same calculation, computed with spherical clouds and varying initial densities. Their initial rotation scales as $L_0 \propto \left(R_\mathrm{cl}/R_\mathrm{H}\right)^2$, whereas the prograde spin-up during their collapse scales as $L_\mathrm{rot} \propto \left(R_\mathrm{cl}/R_\mathrm{H}\right)^5$ (see Sect. \ref{subsect:transfer}). As a result, sparse clouds with $R_\mathrm{cl} \simeq R_\mathrm{H}$ accumulate most of their rotation during their collapse, whereas the rotation of denser clouds remains largely unchanged.}
\label{fig:Rhvar}
\end{figure*}
\begin{subequations}
\begin{align}
    \bar{y}_\mathrm{cl}(t) &= \frac{1}{A_\mathrm{cl}} \int_{r_0-R_\mathrm{cl}}^{r_0+R_\mathrm{cl}}\bar{y}_\mathrm{arc}(r) l_\mathrm{arc}(r) dr \\
    &\simeq r_0 \; \mathrm{sin}\left(\Omega_0 t\right) \nonumber \\ 
    &+ \frac{R_\mathrm{cl}^2}{24 r_0} \left[\left(4 - 9\Omega_0^2 t^2 \right) \mathrm{sin}\left(\Omega_0 t\right) - 9 \Omega_0 t \; \mathrm{cos}\left(\Omega_0 t\right)\right], \label{eq:y_cloud}
\end{align}
\end{subequations}
with the corresponding velocities given by the time derivatives:
\begin{subequations}
\begin{align}
    \bar{v}_\mathrm{x,cl}(t) &\simeq -\Omega_0 r_0\;\mathrm{sin}\left(\Omega_0 t\right)\nonumber \\ 
    &+ \frac{\Omega_0 R_\mathrm{cl}^2}{24 r_0} \left[\left(5+9\Omega_0^2 t^2\right)\mathrm{sin}\left(\Omega_0 t\right)-9\Omega_0 t\; \mathrm{cos}\left(\Omega_0 t\right)\right],
    \label{eq:vx_cloud}
\end{align}
\end{subequations}
\begin{subequations}
\begin{align}
    \bar{v}_\mathrm{y,cl}(t) = 
    &\simeq \Omega_0 r_0\;\mathrm{cos}\left(\Omega_0 t\right) \nonumber \\ 
    &+ \frac{\Omega_0 R_\mathrm{cl}^2}{24 r_0} \left[-\left(5+9\Omega_0^2 t^2\right)\mathrm{cos}\left(\Omega_0 t\right)-9\Omega_0 t\; \mathrm{sin}\left(\Omega_0 t\right)\right]. \label{eq:vy_cloud}
\end{align}
\end{subequations}
These integrals are calculated with Mathematica 12.0 and each have exact solutions in the form of long sums. We provide the Maclaurin series with the variable $\theta_0 \ll 1$ in order to maintain some brevity. To zeroth order, the center-of-mass follows a simple circular motion at distance $r_0$ from the central mass. The strongest deviation is a second-order term that includes a small initial off-set from $r_0$ as well as a time-dependent factor. From Eqs. \ref{eq:x_cloud} and \ref{eq:y_cloud}, we can calculate how the distance of the cloud's center-of-mass ($\bar{r}_\mathrm{cl}$) evolves over time:
\begin{subequations}
\begin{align}
    \delta\bar{r}_\mathrm{cl}(t) &= \sqrt{\bar{x}_\mathrm{cl}^2(t)+\bar{y}_\mathrm{cl}^2(t)} - \sqrt{\bar{x}_\mathrm{cl}^2(t=0)+\bar{y}_\mathrm{cl}^2(t=0)} \\
    &= -\frac{3R_\mathrm{cl}^2 \Omega_0^2 t^2}{8 r_0}. \label{eq:COM_shift}
\end{align}
\label{eq:dr}
\end{subequations}
Interestingly, we find that the center-of-mass in this setup moves inward with a quadratic time dependence. We show the geometrical origin of this change in Fig. \ref{fig:cloudsketch} where we sketch the position of a cloud of particles in Keplerian motion. As the cloud shears out and angular differences increase, its shape deforms from a partial annulus into an increasingly extended single arc. Due to the curvature inherent to circular motion, the center-of-mass of this arc shifts toward the star as a function of time. If the system is evolved for a sufficiently long period of time, the center-of-mass ultimately coincides with the position of the central mass. We can further investigate the trajectory of the cloud's center-of-mass by calculating its semi-major axis ($\bar{a}_\mathrm{cl}$), which follows from the energy balance as:
\begin{equation}
    \bar{a}_\mathrm{cl}(t) = \frac{\Omega_0^2 r_0^3}{2} \left(\frac{\Omega_0^2 r_0^3}{\bar{r}_\mathrm{cl}(t)} - \frac{\bar{v}_\mathrm{x,cl}^2(t) + \bar{v}_\mathrm{y,cl}^2(t)}{2}\right)^{-1}.
\end{equation}
Again, we are mainly interested in its evolution from the initial value as a function of time, so we compute this difference ($\delta\bar{a}_\mathrm{cl}$):
\begin{subequations}
\begin{align}
    \delta\bar{a}_\mathrm{cl}(t) &= \bar{a}_\mathrm{cl}(t) - \bar{a}_\mathrm{cl}(t=0) \\
    &= -\frac{3R_\mathrm{cl}^2 \Omega_0^2 t^2}{2 r_0}. \label{eq:delta_a}
\end{align}
\label{eq:da}
\end{subequations}
In this example calculation, the semi-major axis of the cloud's center-of-mass decreases faster than its distance to the central mass and, hence, its eccentricity increases over time. We have verified that the form of Eq. \ref{eq:delta_a} is unchanged when the shape of the cloud is altered. For example, the pre-factor of 3/2 becomes 1/8 for a cloud in the shape of a closed disk, remains 3/2 for a radial line and tends to zero for a tangential line, whose particles do not experience shear.

\subsection{Transfer of orbital angular momentum to rotation}\label{subsect:transfer}
If an object forms at the center-of-mass of a cloud, its shrinking semi-major axis and increasing eccentricity during the collapse reduce its orbital angular momentum $(\bar{L}_\mathrm{cl})$, which points in the $\hat{z}$-direction in this 2D example:
\begin{subequations}
\begin{align}
    \bar{L}_\mathrm{cl}(t) &= M_\mathrm{cl} \left[\bar{x}_\mathrm{cl}(t) \bar{v}_\mathrm{y,cl}(t) - \bar{y}_\mathrm{cl}(t) \bar{v}_\mathrm{x,cl}(t)\right] \\
    &\simeq M_\mathrm{cl}\Omega_0 r_0^2 \left[1-\frac{1}{24}\left(\frac{R_\mathrm{cl}}{r_0}\right)^2\left(1+18\Omega_0^2t^2\right)\right].
\end{align}
\end{subequations}
While the orbital angular momentum associated with the center-of-mass orbit declines, the total angular momentum of the cloud $(L_\mathrm{tot, cl})$ remains constant over time, equal to:
\begin{subequations}
\begin{align}
    L_\mathrm{tot, cl} &= \int_{r_0 - R_\mathrm{cl}}^{r_0 + R_\mathrm{cl}} \Omega_0 \left(\frac{r}{r_0}\right)^{-\frac{3}{2}} l_\mathrm{arc}(r) \sigma_\mathrm{cl} r^2dr \\
    &\simeq M_\mathrm{cl}\Omega_0 r_0^2 \left[1+\frac{1}{8}\left(\frac{R_\mathrm{cl}}{r_0}\right)^2\right].
\end{align}
\end{subequations}
\begin{figure}[t]
\includegraphics[width=.49\textwidth]{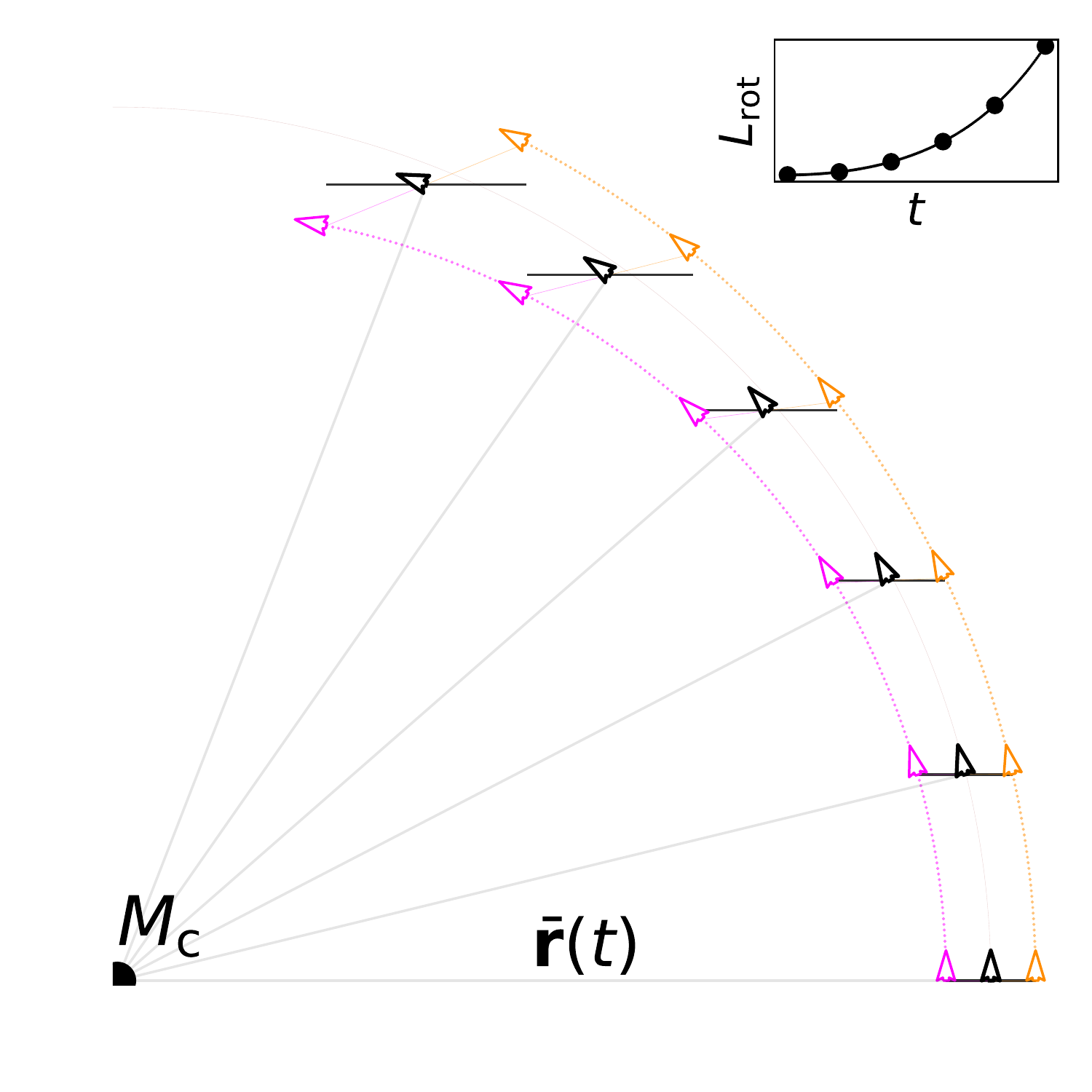}\
\caption{Orbital evolution of two non-interacting particles and their center-of-mass, initiated without rotation. Over time, both the interior (magenta arrow) and the exterior (orange arrow) particles start to revolve around their mutual center-of-mass (black arrow) in a prograde fashion, as shown by the angular displacement that develops relative to the horizontal dashed line. The quadratic prograde spin-up over time is additionally shown in the inserted panel, with the rotation at the different snapshots depicted by the black dots.}
\label{fig:NSG0spin}
\end{figure}
As a thought experiment that represents the simplest model for cloud collapse, we can envision a cloud of material that initially evolves without self-gravity up to time $t$ and then instantly collapses to form an object with position and velocity equal to the cloud's center-of-mass. Conserving total angular momentum, the deficit between the total and orbital angular momentum can be interpreted as this object's rotation and is given by:
\begin{subequations}
\begin{align}
    L_\mathrm{rot, cl}(t) &= L_\mathrm{tot, cl} - \bar{L}_\mathrm{cl}(t) \\
    &\simeq \left(\frac{1}{6} + \frac{3}{4}\Omega_0^2 t^2\right) \left(\frac{R_\mathrm{cl}}{R_\mathrm{H}}\right)^2 L_\mathrm{H},\label{eq:L_rot_analytical}
\end{align}
\label{eq:dL}
\end{subequations}
where we use the Hill radius ($R_\mathrm{H}$) and Hill rotation ($L_\mathrm{H}$) as natural normalization factors in this context (sometimes referred to as Roche units), defined by:
\begin{equation}
    R_\mathrm{H} = r_0 \left(\frac{M_\mathrm{cl}}{3 M_\mathrm{\star}}\right)^\frac{1}{3} \;\;,\;\; L_\mathrm{H} = M_\mathrm{cl} \Omega_0 R_\mathrm{H}^2.
\end{equation}
Eq. \ref{eq:L_rot_analytical} represents the key outcome of our analytical calculation. Again, we have numerically verified that the shape of this key equation is unchanged for different cloud shapes. The pre-factors (1/6, 3/4) become (1/8, 9/16) for a disk, (-1/6, 3/4) for a horizontal line and (1/3, 0) for a vertical line. This agrees with the finding by \citet{Mestel1966} that both a partial annulus and a disk of particles on circular orbits around a point source have prograde spin around their center-of-mass. But more importantly, it shows that the cloud accumulates additional prograde rotational angular momentum over time during its collapse, scaling as $\delta L_\mathrm{rot} \propto t^2$. This rotational angular momentum builds up due to the curvature inherent to orbital motion, which moves the center-of-mass of a cloud toward an increasingly contracted orbit that contains less orbital angular momentum, producing a deficit filled by increased prograde rotation. While we will show in Sect. \ref{sect:ApplCC} that the rate of prograde spin-up depends on the velocity distribution of the cloud, this analytical model is sufficient to discern the main scaling relations, which are as follows. First, we note that the typical timescale to evaluate the prograde spin-up on is the free-fall timescale, which we roughly approximate with the expression for a uniform sphere:
\begin{equation}\label{eq:t_freefall}
    t_\mathrm{ff} = \frac{\pi}{\sqrt{24}} \left(\frac{R_\mathrm{cl}}{R_\mathrm{H}}\right)^\frac{3}{2} \Omega_0^{-1}.
\end{equation}
After substituting this timescale into Eq. \ref{eq:L_rot_analytical}, we find that the rotational build-up during collapse scales as $\delta L_\mathrm{rot} \propto t_\mathrm{ff}^2 \left(R_\mathrm{cl}/R_\mathrm{H}\right)^2 \propto \left(R_\mathrm{cl}/R_\mathrm{H}\right)^5$. The Hill radius can be thought of as the interface between shear and self-gravity, as clouds bigger than $R_\mathrm{H}$ shear out before they collapse. It is really at this interface of shear and self-gravity that the mechanism of rotational gain discussed here is most significant, and its importance rapidly reduces for clouds that begin their collapse at $R_\mathrm{cl}<R_\mathrm{H}$. Second, we note that if the cloud collapses from the Hill radius, the specific rotation of the object that forms, follows a mass scaling of $l_\mathrm{H} = L_\mathrm{H}/ M_\mathrm{cl} \propto \left(M_\mathrm{cl}/M_\mathrm{\star}\right)^{2/3}$, matching the universal rotational trend followed by Solar System asteroids and planets identified by \citet{Goldreich1968}. This proportionality to $M_\mathrm{cl}^{2/3}$ follows the same scaling as rotational predictions from the tidal torque theory that seeks to explain the spin of dark matter halo's and galaxies \citep{Hoyle1951, Peebles1969, Schafer2009}. 
%
%

%

%%%%%%
\section{Numerical evaluation of cloud collapse in orbit}
\label{sect:NumCC}
In this section, we perform numerical simulations of cloud collapse in orbit, using the N-body code REBOUND \citep{ReinLiu2012}. The settings and numerical convergence of the runs are explained in detail in Appendix \ref{appendix:sgcoll}. First, we numerically demonstrate the validity of our main analytical findings. Next, we present a second geometrical argument where the relative position of cloud particles, rather than their shared center-of-mass, is used to illustrate the prograde spin-up during a gravitational collapse. 

\subsection{Comparison with analytical formulation}\label{sect:comparison_analytical}
\begin{figure*}[t] 
\centering
\includegraphics[width=1.0\hsize]{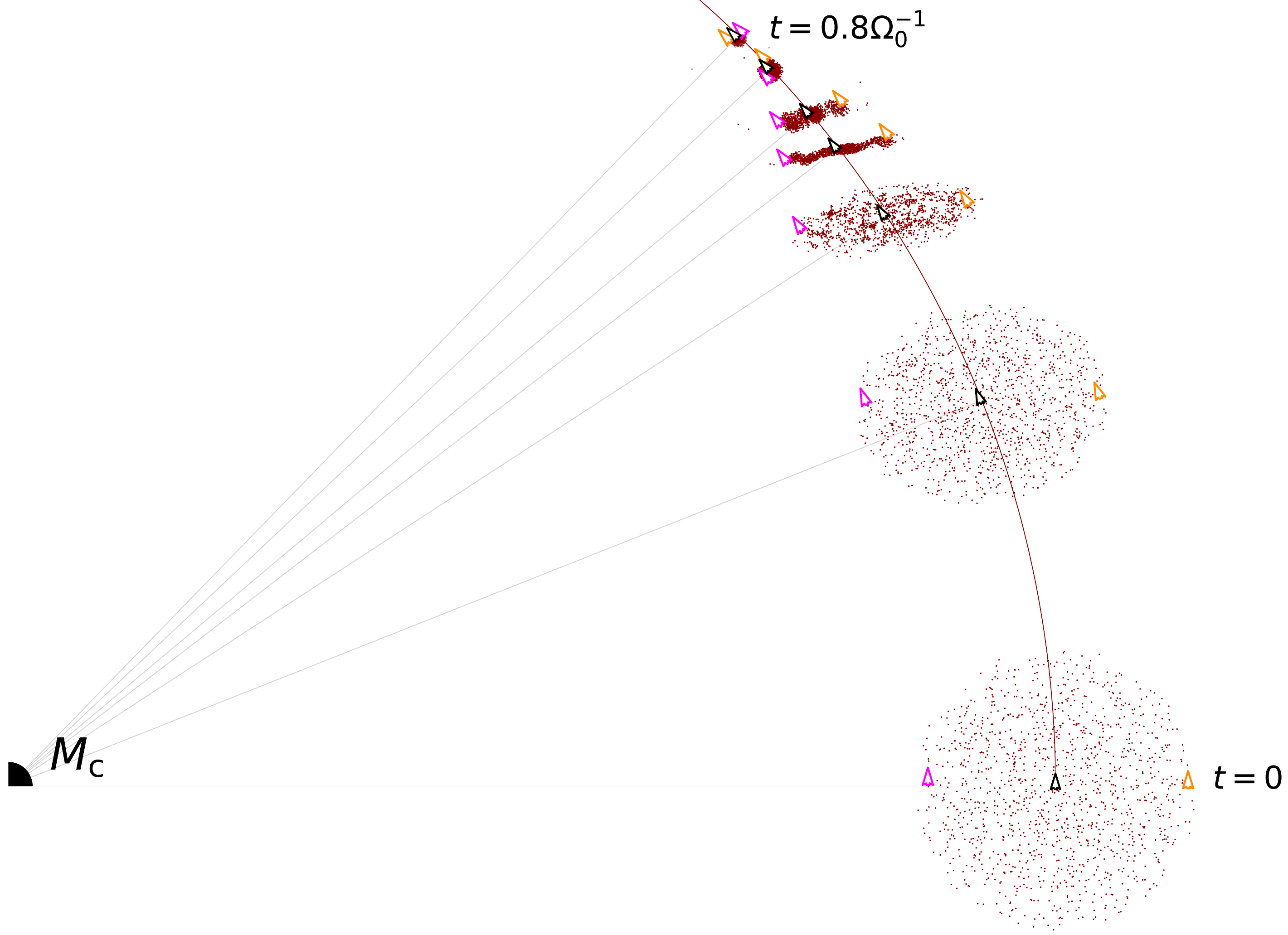}
\caption{Overview of the collapse of a uniform cloud in orbit around a central mass, shown in a stationary frame. The spherical cloud is initiated with size $R_\mathrm{H}$ and without any rotation. The snapshots from t = 0 to $t = 0.8 \; \Omega_0^{-1}$ are enlarged in scale for clarity. We highlight the positions and directions of two particles; one that is initiated closest to the central mass (magenta arrow) and one that starts the furthest away from the central mass (orange arrow). We also indicate the position and direction of the cloud's center-of-mass (black edged arrow) and plot its initial distance (dark red semicircle). At the end of the collapse, the center-of-mass has shifted slightly inward toward the central mass, and the two particles have performed half a prograde revolution, confirming the prograde rotation of the object that is formed.}
\label{fig:collpanelshelio}
\end{figure*}
The main simplification of our analytical example, besides its fixed 2D geometry, is that it approximates the prograde spin-up without accounting for the orbital changes induced by self-gravity. In the left panel of Fig. \ref{fig:Rhvar}, we numerically compute the same example including self-gravity to illustrate the spin-up of a cloud that actually collapses. The two curves initially follow the same trend of $\delta L_\mathrm{rot} \propto t^2$ during the early phase of the collapse. Then, the spin-up of the collapsing cloud begins to level off as it contracts sufficiently. The process of prograde spin-up eventually halts entirely when the cloud becomes a rigid body, and it can no longer shear out. Nevertheless, the analytical example provides a decent approximation when the spin-up is evaluated at $t_\mathrm{ff}$.

Our analytical arguments predict that prograde spin-up scales as $\delta L_\mathrm{rot} \propto \left(R_\mathrm{cl}/R_\mathrm{H}\right)^5$, more steeply than the cloud's initial rotation, which scales as $L_0 \propto \left(R_\mathrm{cl}/R_\mathrm{H}\right)^2$ (see Eqs. \ref{eq:L_rot_analytical}, \ref{eq:t_freefall}). To test this trend and see if it holds for differently shaped clouds, we perform numerical simulations of uniform, spherical clouds with different initial sizes, again initialized on non-eccentric, Keplerian orbits. We show the prograde spin-up of these runs in the right panel of Fig. \ref{fig:Rhvar} for initial cloud sizes between $0.5-1 R_\mathrm{H}$. As analytically predicted, the spin-up can exceed the initial rotation in scale, but only when the cloud nearly fills its Hill radius prior to collapse. Indeed, the fraction of accumulated rotation drops from $\delta L_\mathrm{rot}/L_0 =1.40$ to $\delta L_\mathrm{rot}/L_0 =0.16$ when the cloud's initial size is halved. This factor ~9 decline is similar to the factor 8 predicted analytically and re-iterates our finding that, while the rotation of dense ($R_\mathrm{cl} \ll R_\mathrm{H}$) clouds remains largely unchanged during their collapse, clouds that fill their Hill sphere prior to collapse gain a substantial prograde component to their spins.

\subsection{Prograde spin-up in the absence of initial rotation}
It is also possible to visualize the prograde spin-up during a gravitational collapse in orbit by tracking the relative motion of a cloud's inner- and outermost particles. A similar analysis was performed by \citet{Artemev1965} to show that a circular patch of material in Keplerian motion has prograde spin, a fact that also follows from the positive vorticity. We expand upon it here to illustrate the prograde spin-up of a cloud over time.

\label{sect:collapse_panels_global}
\subsubsection{2-particle dynamics without self-gravity}
The prograde spin-up during collapse is most striking in clouds that have no initial rotation, which we ensure by initiating all particles with (vector) velocities equal to that of their shared center-of-mass (see Appendix \ref{Appendix:rotation_IC}). Prior to showing the rotational build-up in an N-body collapse, we first illustrate the underlying process in Fig. \ref{fig:NSG0spin}, where we consider the orbital motion of two horizontally aligned, non-interacting particles in a stationary frame. Their shared center-of-mass coordinate is shown with a black arrow and is initiated with a non-eccentric Keplerian velocity.

Rather than focusing on the inward shift of the center-of-mass shift due to shear, which is not always visible relative to the cloud's scale, we focus here on the relative motion of the particles. The innermost particle is more tightly bound to the central mass and traces out an elliptical orbit that is characterized by a greater angular velocity ($\dot{\theta}$) than that of the outermost particle. As a result of the orbital geometry, where its y-coordinate is modulated by a factor $\mathrm{sin}(\theta)$ in this stationary frame, the inner particle begins to drop below the exterior one. Consequently, it is seen to effectively wrap below their shared center-of-mass in a prograde fashion. This behavior is mirrored in reverse by the outer particle, whose angular velocity is lower and, therefore, wraps around their center-of-mass from the top, again producing a prograde rotation. We emphasize this relative movement visually in Fig. \ref{fig:NSG0spin} with a horizontal dashed line, drawn through the center-of-mass. In the inserted panel, the rotation relative to the center-of-mass is seen to rise quadratically over time, producing the same trend of prograde spin-up as a derived in Sect. \ref{sect:analytical}.

\subsubsection{N-body collapse of a spherical cloud}
We now perform a cloud collapse, using the same initial condition without rotation, but this time for a uniformly packed cloud with size $R_\mathrm{H}$ and with self-gravity and collisions enabled in REBOUND. We visualize the evolution of the collapsing cloud in Fig. \ref{fig:collpanelshelio}\footnote{Fig. \ref{fig:collpanelshelio} shows the collapse in a stationary frame that is centered on a much more massive central mass. This poses the visual difficulty that any collapsing cloud is tiny in comparison to the scale of its orbit. To circumvent this, we place sub-panels at the corresponding center-of-mass positions where we zoom in. The cloud scale is, therefore, exaggerated for clarity but the collapse evolution is still visualized in a stationary frame.}, again shown in a stationary frame. If we compare the stages of collapse over time with Fig. \ref{fig:NSG0spin}, we observe that the general behavior remains unchanged. The difference is caused by the addition of self-gravity, which draws all particles toward their shared center-of-mass over time. In the intermediate stages of the collapse, this self-gravity reshapes the cloud into an aligned bar, as is the case when any rotating spheroid collapses \citet{Lin1965}. Just like in the two-particle example, the interior particle drops below the center-of-mass - while those exterior remain above it. As the cloud contracts further and self-gravity intensifies, this prograde wrapping of particles on either side of the center-of-mass translates into a real prograde rotation of the object that forms.

\section{Application: The streaming instability}\label{sect:SI}
We highlight the streaming instability as a small-scale example of a scenario where particle clouds form and then collapse due to their self-gravity while subject to a strong external force, in this case gravity from a central star. Streaming instabilities occur when small over-densities in the proto-planetary disk locally accelerate the gas, producing dense filaments where small solids concentrate and collapse to form planetesimals \citep{youdingoodman2005, JohansenEtal2007, JohansenEtal2009}. Hydrodynamic simulations of the streaming instability in the proto-Kuiper belt have shown that it is possible to create equal-mass binaries in this manner \citep{Nesvorny2010}, which naturally seem to form with a strong preference for prograde rotation \citep{Nesvorny2019, Nesvorny2021}. These simulations provide a striking agreement with the observed abundance of prograde, equal-mass binaries in the dynamically cold class of Kuiper belt objects \citep{Noll2008, Fraser2017, Grundy2019}. 

In this section, we evaluate whether the mechanism of prograde spin-up can drive this bias in rotational direction. Similar to \citet{Robinson2020}, we model the final collapse stage of the streaming instability with N-body simulations, and do not investigate the hydrodynamic onset of the instability itself, which sets the initial conditions for the collapse. Instead, we take a range of representative initial conditions and calculate the rotational gain in each of these scenarios. The effect of gas drag in this final collapse phase can be neglected for our purposes here, assuming that the cloud has a dust-to-gas ratio exceeding unity \citep{Nesvorny2010}.

\begin{figure} 
\includegraphics[width=.49\textwidth]{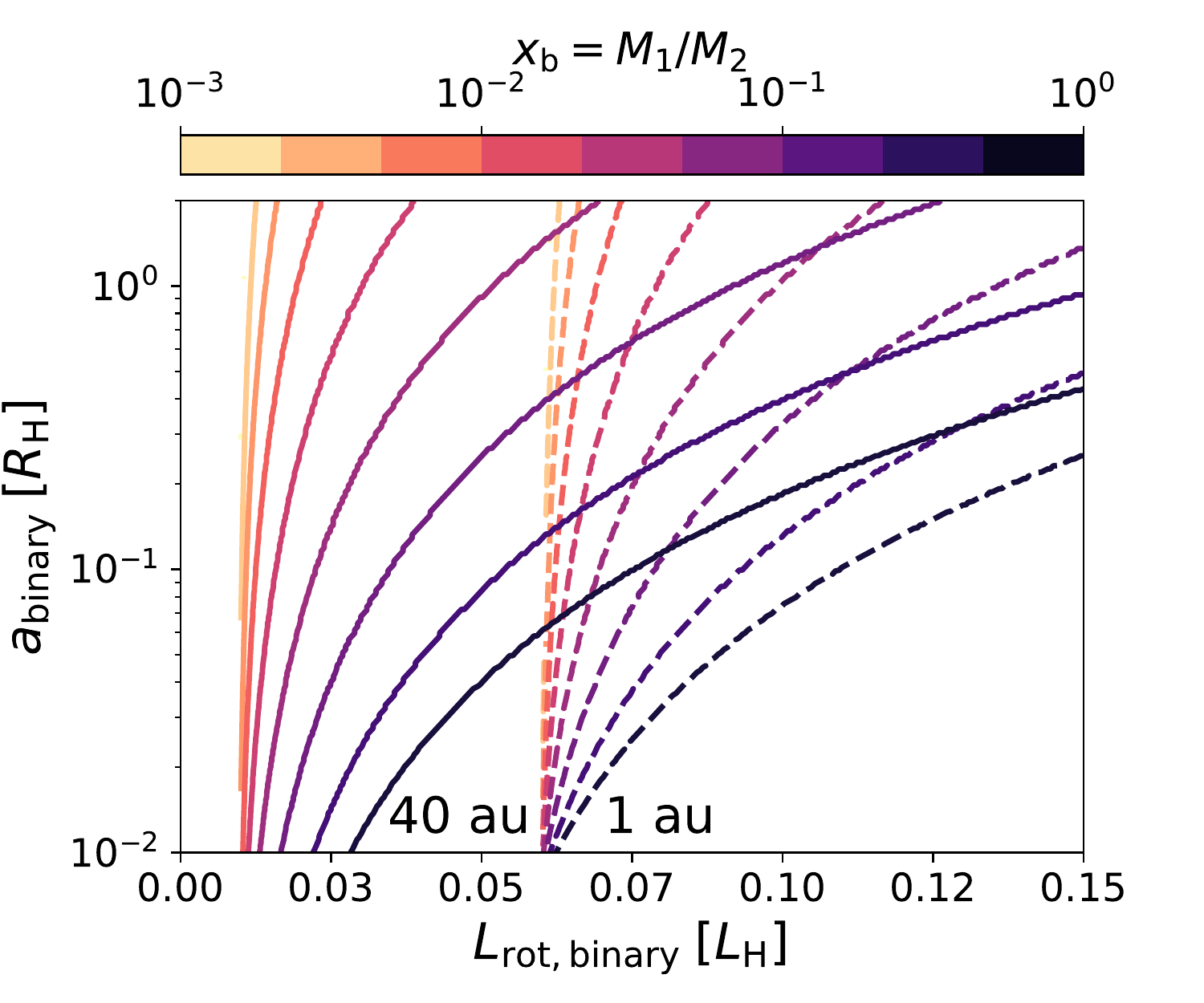}\
\caption{Possible binary configurations for a given rotational angular momentum budget ($L_\mathrm{rot,binary}$) at two distances from the central star (40 au, solid; 1 au, dashed), assuming that the components themselves spin at near-breakup rates with density $\rho_\mathrm{p}=1$. When $L_\mathrm{rot,binary} <0.06 \; L_\mathrm{H}$, only single planetesimals are formed at 1 au as the break-up spin exceeds the angular momentum budget, whereas this critical value lies between 0.02-0.03 $L_\mathrm{H}$ at 40 au. Large amounts of rotational angular momentum $\gtrsim 0.1 L_\mathrm{H}$ can only be contained in binaries with increasingly wide mutual orbits and increasingly similar masses, especially in the outer disk.}
\label{fig:possible_binaries}
\end{figure}
\begin{figure*}[t] 
\centering
\includegraphics[width=\hsize]{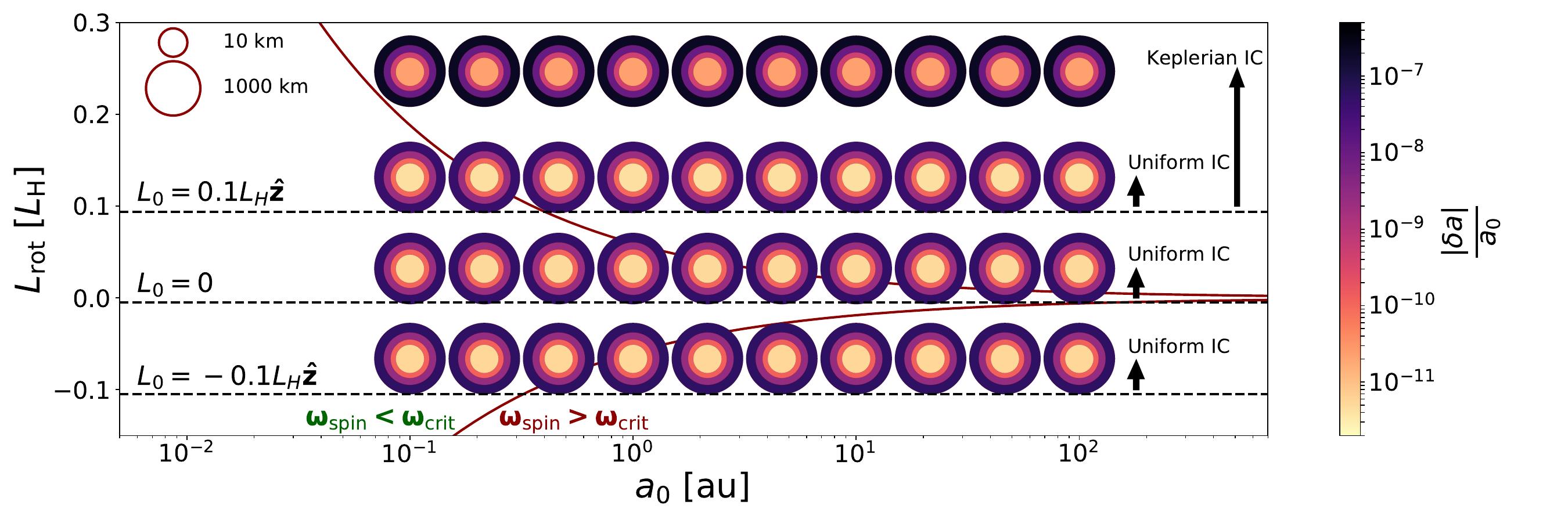}
\caption{Prograde spin-up in the collapse phase of the streaming instability, visualized for a wide range of orbital separations, asteroid sizes and initial conditions. The dashed horizontal lines tangent to the circles represent the rotation prior to collapse, while the arrows indicate the magnitude of prograde spin-up. The ringed circles represent the size of the asteroid that forms, and their colors show the total center-of-mass displacement during the collapse $\delta a$. The systems within the dark red lines represent cases where there is insufficient rotational angular momentum to form binaries ($\omega_\mathrm{spin}<\omega_\mathrm{crit}$). The systems above and below the red lines show the formation of prograde and retrograde binaries, respectively. The preference for prograde binary formation is clearly visible, especially in the outer disk.}
\label{fig:binary_collapse}
\end{figure*}

\subsection{Binary formation criterion}\label{sect:ApplCC}
We first examine how much rotational angular momentum is required to form a binary. If a spherical asteroid has a uniform density $\rho_\mathrm{p}$, its moment of inertia for rotation is $I_\mathrm{p} = (2/5)M_\mathrm{p}R_\mathrm{p}^2$. If the asteroid's material strength is neglected - a reasonable assumption based on their low typical internal strengths \citep[e.g.][]{Burns1975, Degewij1976, Carbognani2017, Persson2021} - its spin can at most reach the breakup limit of $\omega_\mathrm{crit} = \sqrt{4\pi G \rho_\mathrm{p} / 3}$ \citep{Pravec2000}. The amount of angular momentum contained in a single asteroid's spin can then be written as a fraction of this limit:
\begin{subequations}
\begin{align}
    L_\mathrm{spin,p} &= I_\mathrm{p} \omega_\mathrm{spin} \\
    &\simeq 0.06 \; L_\mathrm{H} \left(\frac{d}{\mathrm{AU}}\right)^{-\frac{1}{2}} \left(\frac{M_\star}{\mathrm{M_\odot}}\right)^\frac{1}{6} \left(\frac{\rho_\mathrm{p}}{1}\right)^{-\frac{1}{6}} \frac{\omega_\mathrm{spin}}{\omega_\mathrm{crit}},
\end{align}
\end{subequations}
where $\omega_\mathrm{spin} \leq \omega_\mathrm{crit}$. Any excess rotational angular momentum can be accounted for by ejecting a portion of the mass, or by forming a binary (with mass ratio $x_\mathrm{b} = M_1/M_2$ and separation $a_\mathrm{b}$), rather than a single planetesimal. In the idealized case without mass ejection, the complete rotational angular momentum budget gets divided into two spin terms around the respective asteroid centers (that sum to $L_\mathrm{spin,p1} + L_\mathrm{spin,p2} = L_\mathrm{spin,binary}$), as well as an orbital term for the motion of the binary around their shared center-of-mass ($L_\mathrm{orb,binary}$). The spin component follows from the moments of inertia as:
\begin{equation}\label{eq:l_spin}
    L_\mathrm{spin,binary}=L_\mathrm{spin,p}\left(1+x_\mathrm{b}\right)^{-\frac{5}{3}}\left(1+x_\mathrm{b}^\frac{5}{3}\right),
\end{equation}
while the orbital rotation around the binary's center-of-mass can be calculated from their angular velocity ($\omega_\mathrm{b} = \sqrt{GM_\mathrm{p}/a_\mathrm{b}^2}$):
\begin{subequations}
\begin{align}
    L_\mathrm{orb,binary} &= M_1 \omega_\mathrm{b} a_1^2 + M_2 \omega_\mathrm{b} a_2^2 \\
    &= L_\mathrm{H} \frac{x_\mathrm{b}}{(1+x_\mathrm{b})^2} \left(\frac{a_\mathrm{b}}{R_\mathrm{H}}\right)^{\frac{1}{2}}. \label{eq:xb_exact}
\end{align}
\end{subequations}
Assuming that the collapse forms a bound system of at most two objects, the sum of Eqs. \ref{eq:l_spin} and \ref{eq:xb_exact} specifies the possible binary configurations ($x_\mathrm{b}, a_\mathrm{b}$) for a given budget of rotational angular momentum. We examine these configurations in Fig. \ref{fig:possible_binaries} at the location of the Kuiper belt (40 au) and at the inner edge of the main belt (1 au). The objects in the binaries are assumed to spin at breakup rate, in agreement with most Solar System asteroids \citep[e.g.][]{Carbognani2017, Persson2021}.

The first thing to note is that the formation of binaries is more likely at greater distances from the central star, where more angular momentum is available. This is due to the fact that the total rotational angular momentum generated in a collapse is constant across the disk in Hill units, which represent increasing angular momentum at greater orbital separations ($L_\mathrm{H} \propto r_0$). Furthermore, we note that the more equal the mass of the binary components, the more rotational angular momentum the combined system can store. When binaries are formed, therefore, they more commonly form with equal mass ratio's in the outer disk. This trend is indeed reflected in the Solar System's asteroid/comet population. In the next subsection, we take a closer look at the magnitude of the rotational accumulation with different initial conditions for the collapsing cloud.

\subsection{Dependence of spin-up on initial conditions}
The magnitude of the prograde spin-up during a gravitational collapse depends on the initial conditions of the cloud, including its size (Sect. \ref{sect:comparison_analytical}), velocity distribution and distance from the central star. We now examine the importance of the latter two variables by running the following array of N-body collapses:

\begin{enumerate}
    \item We vary the semi-major axis between $a_0 \in [10^{-1}, 10^{2}] \; \mathrm{au}$ and vary the cloud's mass between the equivalent of planetesimals with radii $R_\mathrm{p} \in [10, 10^{3}] \; \mathrm{km}$, assuming $R_\mathrm{cl}=R_\mathrm{H}$.
    \item We run this grid for four different rotational initial conditions, described in detail in Appendix \ref{Appendix:rotation_IC}:
    \begin{enumerate}
        \item \textit{Keplerian} initial conditions, where cloud particles are initiated with non-eccentric local Keplerian velocities, yielding an initial rotational angular momentum of $L_0 \simeq 0.1 L_\mathrm{H}$.
        \item \textit{Zero-rotation} clouds that are initiated with velocities equal to that of the center-of-mass.
        \item \textit{Uniform} initial prograde/retrograde rotation, where clouds rotate at a constant angular rate with a magnitude $L_0 = \pm 0.1L_\mathrm{H}$. This value is chosen to be identical in magnitude to the Keplerian setup and is similar to the log-median rotational angular momentum of bound clumps in the hydrodynamic simulations of \citet{Nesvorny2021}.
    \end{enumerate}
\end{enumerate}
We show the results in Fig. \ref{fig:binary_collapse}, where we indicate the prograde spin-up as well as the corresponding inward shift in the semi-major axis. Consistent with our analytical prediction, the rotational build-up is both scale-invariant and distance-invariant when expressed in Hill units. As predicted, the reduction of the cloud's semi-major axis during the collapse scales positively with the cloud mass and is proportional to its distance to the central star.

Interestingly, the initial distribution of velocities in the cloud makes a large difference in the magnitude of the rotational gain. Whereas a cloud that begins with circular, Keplerian velocities attains an additional prograde rotation of $\delta L_\mathrm{rot}\simeq 0.15 L_\mathrm{H}$, clouds that begin with uniform rotation only gained $\delta L_\mathrm{rot}\simeq 0.05 L_\mathrm{H}$. This difference can be explained by the fact that prograde spin-up is driven by shear between the particles. If the initial conditions reflect more uniform motion, the cloud shears out to a less extended arc during the collapse and the rotational gain is reduced. Nevertheless, even clouds that start without any rotation accumulate a substantial prograde value of $L_\mathrm{rot}\simeq 0.05 L_\mathrm{H}$, which is enough to trigger unequal-mass binary formation outside a few au. In the case of Keplerian initial conditions, the rotational gain (to total $L_\mathrm{rot}\simeq 0.25 L_\mathrm{H}$) is enough to form equal-mass binaries across the proto-planetary disk (See Fig. \ref{fig:possible_binaries}).

%%%%%%
\section{Discussion} \label{sect:Discussion}
The mechanism of prograde spin-up is generally applicable to the gravitational contraction of any clouds that are subject to an external, central force. In any context, the key observational signature of this mechanism is a preferential alignment between the spin and orbital vectors of the objects that form. However, the rotational gain scales steeply as $\delta L_\mathrm{rot} \propto (R_\mathrm{cl}/R_\mathrm{H})^5$, such that only clouds whose sizes are set by the interplay of tidal shear and self-gravity ($R_\mathrm{cl} \sim R_\mathrm{H}$) experience significant spin-up when they collapse. Expressed in terms of a spherically uniform density, the pre-collapse system should not be much denser than $\rho_\mathrm{H} = 9M_\mathrm{C}/(4\pi r_0^3)$, offering a scale-dependent benchmark. In this section, we will first compare our simple simulations of binary asteroid formation by streaming instability to other works. Next, we move up on the scale ladder and discuss the potential implications of prograde spin-up for the rotation of planets, stars and molecular clouds.

\subsection{Comparison to hydrodynamic simulations of binary formation in the Kuiper belt}
We presented a simple application of prograde spin-up to the streaming instability, where we used a set of idealized initial conditions to show how perturbations by Solar gravity provide enough rotation to form prograde binary systems. In reality, such clouds have morphologies and velocity distributions that are set in a more complex manner by interactions between the nebular gas and the solids, including back-reactions \citep{youdingoodman2005, JohansenEtal2007, JohansenEtal2009}. These processes were accounted for in more detail in the works by \citet{Nesvorny2019, Nesvorny2021}, who simulated the onset of the instability and initial stages of collapse using the hydrodynamics code ATHENA \citep{Stone2008}, prior to simulating high-resolution binary formation in the final collapse stage with the N-body code PKDGRAV \citep{Stadel2001}. Notably, \citet{Nesvorny2019, Nesvorny2021} exclude stellar gravity in their PKGRAV simulations, but include it in their ATHENA runs, allowing their results to be influenced by prograde spin-up during the key early collapse stage. A comparison with their results illustrates two points: First, the broad distribution of scaled angular momenta that is generated by their ATHENA simulations, both in magnitude and direction, emphasizes the importance of modeling the onset of the streaming instability when considering the initial conditions of the collapse. Second, the fact that our simple model replicates their key result, namely the tendency of streaming instabilities to form prograde binaries, indicates that the mechanism of prograde spin-up may well drive the rotational outcome in both models. Indeed, this suggestion is reinforced by the data shown in the supplementary Fig. 4 of \citet{Nesvorny2019}, which indicates that the colatitude distribution of the vorticity vectors in their simulations is initially broad, and that the prograde bias only appears when the bound clumps collapse to form binary systems. We suggest that the relative importance of prograde spin-up to the turbulence of the instability determines the amount of prograde bias. 

Rather than investigate the precise morphologies of these binary systems that form in SI, as was done with high-resolution studies by \citet{Robinson2020} and \citet{ Nesvorny2021}, we studied the mechanism behind the rotational evolution of material during the collapse. Conveniently, the prograde spin-up is largely independent of the chosen resolution or physical parameters like the coefficient of restitution and the particle size (See Appendix \ref{appendix:sgcoll}). We should note, however, that the possibility of material ejection during the late stages of collapse means that not all the rotational angular momentum built up by the material necessarily ends up in the binary. The details tend to be more sensitive to parameter variations \citet{Robinson2020,Nesvorny2021}, and high-resolution simulations remain necessary to study the precise configurations that can form.

\subsection{Relevance of prograde spin-up during planet formation}
At the scale of planets, the mechanism of prograde spin-up is most clearly applicable to the formation of giant planets by gravitational instability \citep[e.g.][]{Kuiper1951, Cameron1978, Boss1998}. The spins of these large planets on wide orbits are only just beginning to be uncovered, with the only two currently known cases pointing to moderate or large obliquities \citep{Bryan2020, Bryan2021}. Recent SPH and hydrodynamical simulations of gas giant formation, which are sensitive to prograde spin-up, indicate that gravito-turbulent discs can yield obliquities up to 90 degrees \citep{Jennings2021}, although most objects seem form with more alignment \citep{Hall2017}. One notable difference with asteroid formation is that a collapse that forms a gas giant is halted by gas pressure, rather than collisions between solids. In both cases, however, these arresting forces become relevant only when the collapse is almost complete and significant prograde spin-up has already taken place. Alternatively, in the framework of core accretion \citep[e.g.][]{Mizuno1980}, gas giants could have accreted most of their mass and prograde angular momentum during the stage of runaway gas accretion \citep{Machida2008, Dittmann2021}. In this case too, the nebular gas accretes onto the planet from a range of non-eccentric orbits while exposed to stellar gravity, allowing for prograde spin-up to increase planetary rotation. Further work is required to resolve the statistical spin distribution of this class of planets and to investigate the role that prograde spin-up plays in setting their rotation.

The relevance of prograde spin-up to the formation of terrestrial planets is yet more difficult to assess. While their embryo's may have formed in streaming instabilities, terrestrial planets likely accreted most of their mass and rotation from nearby planetesimals \citep[e.g.][]{Pollack1996}, from the aerodynamic capture of pebbles \citep{Ormel2010, Lambrechts2012}, or possibly from late giant impacts \citep[e.g.][]{Wetherill1985, Canup2012}. Seen from a rotational perspective, however, none of these scenarios seem to offer satisfactory explanations for the scale or prograde bias of planetary spins. The accretion of planetesimals fails to deliver sufficient rotation \citep{LissauerKari1991, DonesTremaine1993} while stochastic, giant impacts lead to isotropic spin distributions \citep{Safronov1966, DonesTremaine1993i}. Pebble accretion has been found to produce large, prograde rotation in some regimes \citep{Johansen2010, Visseretal2020}, although this may be dampened when proto-planetary envelopes are accounted for. There is, however, an alternative formation scenario that does involve a terrestrial-scale gravitational collapse. In the inside-out formation channel suggested by \citet{Chatterjee2014, Hu2016, Hu2018, Mohanty2018, Cai2022}, terrestrial cores are proposed to form out of dense pebble traps at the edges of dead zones. If planet formation indeed sometimes proceeds through this channel, the rotation of the planets it produces will likely be affected by prograde spin-up.

\subsection{Prograde spin-up of molecular clouds and stars}
Star formation is a hierarchical process that involves collapse on different scales, with potential applications for prograde spin-up at different levels. The process begins with the formation of a molecular cloud, the largest of which are observed to be concentrated toward galactic spiral arms \citep{Lee2001,Blitz2005,Stark2005}, where they form out of the interstellar medium, likely via gravitational instability driven by both stars and gas \citep{Ballesteros-Paredes2007}. At this scale, prograde spin-up is driven by the gravitational potential of the galaxy and contributes to an alignment between the rotation of the molecular cloud and its galactic orbit. Observations of molecular clouds in M33 and M51 confirm this trend, with strong measured prograde biases in their rotation \citep{Braine2018, Braine2020}. It is not clear what fraction of their rotation can be attributed to prograde spin-up, though, as both M33 and M51 have rising velocity curves that naturally lead to prograde rotation when a section of the disk contracts. Interestingly, \citet{Jeffreson2020} performed numerical simulations of molecular cloud formation in the dynamical galactic environment and found that the prograde bias remains even when the galactic rotation curve is declining and shear is expected to yield retrograde orbits. This persisting prograde bias fits the mechanism of prograde spin-up, but more work is required to investigate a potential connection.

Moving down in scale, most stars are known to form in dense, gravitationally bound sub-structures within molecular clouds \citep[e.g.][]{McKee2007, Lee2012, Lee2016b}, often referred to as star-forming clumps. When these clumps collapse and fragment, they form proto-clusters that often contain many thousands of stars \citep[e.g.][]{McKee2007, Lee2012, Lee2016b}.
For the stars that form, the gravity from the star-forming clump itself acts as the external, central force necessary for prograde spin-up, which will act to align the angular momentum of the stellar spin (or binary orbit) with any orbital angular momentum within the cluster. While the available angular momentum is expected to reduce during star formation via magnetic breaking, the sign of the spin is maintained. If a star-forming clump contains systematic rotation, a spin-orbit alignment will translate to a correlation between the stellar spins in the cluster. Until recently, no evidence of such a correlation was known, and the first spectroscopic studies of the
young, low-mass open clusters Pleiades and Alpha Per were observed to be isotropic \citep{Jackson2010, Jackson2018}. Recently, however, new astro-seismology techniques has shown that stars in the large open clusters NGC 6791 and NGC 6819 possess a strong inter-cluster spin alignment \citep{Corsaroetal2017}, pointing to an imprint of spin-orbit alignment during their formation. Hydrodynamic simulations indicate that this spin-alignment can indeed arise when the star-forming clumps are mostly rotationally supported prior to their global collapse \citep{Lee2016a, Corsaroetal2017}, though further numerical simulations that track the rotational evolution during the key early collapse of the star-forming clump are required to assess the relative importance that prograde spin-up plays in this process.

%

%%%%%%
\section{Conclusions} \label{sect:Conclusions}
Larger asteroids, planets, stars in some clusters, as well as molecular clouds seem to possess a preferential alignment between their spins and orbital vectors, hinting at a shared link to their parent structures. In this work, we describe a process that can cause spin-orbit alignment based on the gravitational collapse of clouds in orbit around an external, central potential. We perform illustrative analytical and N-body calculations to show that a cloud's rotational angular momentum relative to its center-of-mass is not conserved when it contracts next to such a potential, even if no material is ejected. Instead, because particles in a cloud shear away from one another over time - and do so on curved paths - their combined center-of-mass moves inward toward the source of the potential (See Fig. \ref{fig:cloudsketch}). The orbital angular momentum that is thus liberated, adds a prograde component to the spin of the object that forms. Equivalently, the process of spin-up can be understood from the prograde wrapping of inner and outer particles around the cloud's center-of-mass, which we visualize in both a stationary (Figs. \ref{fig:NSG0spin}, \ref{fig:collpanelshelio}) and co-rotating (Fig. \ref{fig:collpanels}) frame. The basic properties of the prograde spin-up mechanism are as follows:
\begin{enumerate}
    \item Clouds that orbit around external, central potentials build up prograde rotation when they collapse due to self-gravity. The spin-up develops quadratically over time ($\delta L_\mathrm{rot}/L_\mathrm{H} \propto t^2$), before slowing down when the collapse completes ($t\sim t_\mathrm{ff}$).
    
    \item We find that the total rotational gain scales as $\delta L_\mathrm{rot}/L_\mathrm{H} \propto t_\mathrm{ff}^2 \propto \left(R_\mathrm{cl}/R_\mathrm{H}\right)^5$. The fifth-order scaling means that prograde spin-up is more important for clouds with lower densities prior to their collapse.
    
    \item If the external potential is generated by a point mass and the initial cloud is spherical, we find a total increase in rotational angular momentum around $\delta L_\mathrm{rot} \simeq 0.05-0.15 \; M_\mathrm{cl}\Omega_0 R_\mathrm{cl}^2$. Out of the studied initial conditions, the magnitude of prograde spin-up is greatest when the cloud's particles orbit with circular velocities prior to the collapse, allowing for efficient tidal shearing.
\end{enumerate}
When applied to the Solar System, we suggest that this mechanism of prograde spin-up provides an explanation for the observed spin-orbit alignment of trans-Neptunian binaries \citep{Grundy2019}, provided that they formed in streaming instabilities \citep{youdingoodman2005, JohansenEtal2007, JohansenEtal2009}. As such, prograde spin-up could operate to drive the recent simulation results of \citet{Nesvorny2019, Nesvorny2021}, who recently replicated this preferential spin-orbit alignment. The relevance of prograde spin-up to the formation of objects by gravitational collapse on larger astrophysical scales remains open for further investigation. It is likely, however, that the universal applicability of prograde spin-up contributes to the ubiquity of spin-orbit alignment on different scales. Compared to the rotation contained in shear: $L_\mathrm{rot} / L_\mathrm{H} \propto \left(R_\mathrm{cl}/R_\mathrm{H}\right)^{2}$, prograde spin-up becomes important when the size of the cloud prior to collapse is comparable to the Hill radius. For objects that form in this interface between self-gravity and shear, prograde spin-up naturally produces a spin-orbit alignment, even in an environment of retrograde shear.

\section*{Acknowledgements}
\tiny{We are deeply indebted to Carsten Dominik and Amy Bonsor for their support and mentorship. We thank Chris Ormel and Helong Huang for providing comments and corrections on the manuscript for this work. In addition, we are grateful for useful discussions on this topic with Mark Wyatt, Cathie Clarke, Elliot Lynch, Tess Roovers, Sjoerd van der Heijden, Mitchell Yzer, Tom Konijn, and Alexander Hackett. We thank the anonymous referee for an encouraging report. RGV acknowledges funding from the Dutch Research Council, ALWGO/15-01. MGB acknowledges the support of a Royal Society Studentship, RG 16050.}
\bibliographystyle{aa}
\bibliography{rotation} % please keep alphabetic

\appendix 
\normalsize{
%

%%%%%%
\section{Rebound settings and convergence}
\label{appendix:sgcoll}
In this appendix, we describe the settings of our REBOUND runs used to model cloud collapse in the final stage of the streaming instability. First, we spread the particles evenly across the clouds to ensure that their density is uniform. To do so, we employ a Voronoi \citep{Voronoi1908} algorithm with Lloyd iterations \citep{,Lloyd1982} until a high level of uniformity is achieved.
The N-body integration itself is done with the variable step solver IAS15 \citep{ReinSpiegel2015}.

As the clouds contract, particles begin to collide. We model these collisions with an inflated particle approach to reduce the number of numerical particles ($N_\mathrm{n}$) from the number of physical particles ($N_\mathrm{p}$). In this approach, the radius of the numerical particles ($s_\mathrm{n}$) is blown up to maintain total cross-section: $s_\mathrm{n} = s_\mathrm{p} \left(N_\mathrm{p} / N_\mathrm{n}\right)^{1/2}$. An interesting convenience that follows from the larger particles, as pointed out by \citet{Reinetal2010}, is that the use of a gravitational softening parameter becomes redundant. We note that the inflated particle approach is only valid if the initial volume filling factor of the particles is well below the size of the cloud prior to collapse, in our simulations typically set to the Hill radius. Because the Hill radius is proportional to the orbital separation, the filling factor of a cloud with otherwise identical parameters ($N_\mathrm{n}, s_\mathrm{p}$) is reduced when it is positioned further from the star. In order to ensure that the cloud collapse behaves in the same manner at every semi-major axis, we instead fix the initial volume filling factor at $f = 10^{-3}$ in every simulation. This stylized approach is not suitable for resolving the physical system that forms from a collapse, as was modeled by \citet{Robinson2020,Nesvorny2021}, but it is suitable for gauging the total amount of spin-up during the collapse. It allows for the use of far fewer numerical particles while maintaining convergence in this result. We show this in Fig. \ref{fig:convergenceFill}, where the obtained rotation is indeed not dependent on the chosen number of numerical particles nor on the choice of filling factor. The coefficient of restitution in our model is taken to be zero, a choice that was not found to influence the results.

\begin{figure}[t]
     \includegraphics[width=.49\textwidth]{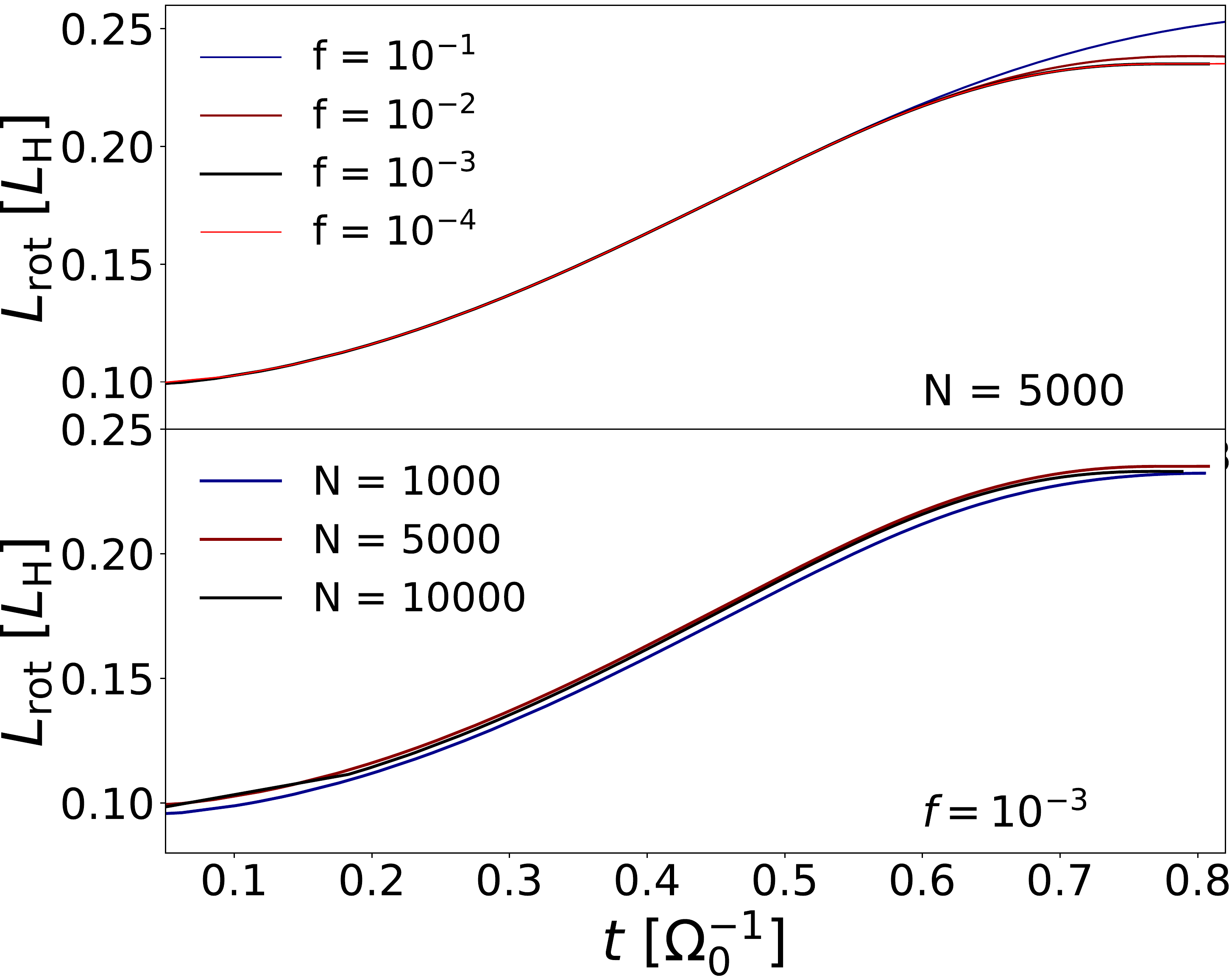}
     \caption{Dependence of the prograde spin-up over time on numerical parameters. Top: variation of the filling factor of the simulation box between $10^{-4} \leq f \leq 10^{-1}$. The runs correspond to $N_\mathrm{n} = 5000$ numerical particles and are run at 5 AU. Bottom: variation of the number of numerical particles $10^{3} \leq N_\mathrm{n} \leq 10^4$, computed with a filling factor of $f=10^{-3}$. The magnitude of prograde spin-up is found to be insensitive to these numerical parameters.}
     \label{fig:convergenceFill}
\end{figure}
%
%%%%%%
\section{Angular momentum transformation between local rotating and inertial frames}\label{appendix:transformation}
\begin{figure*}[t!] 
\centering
\includegraphics[width=\textwidth]{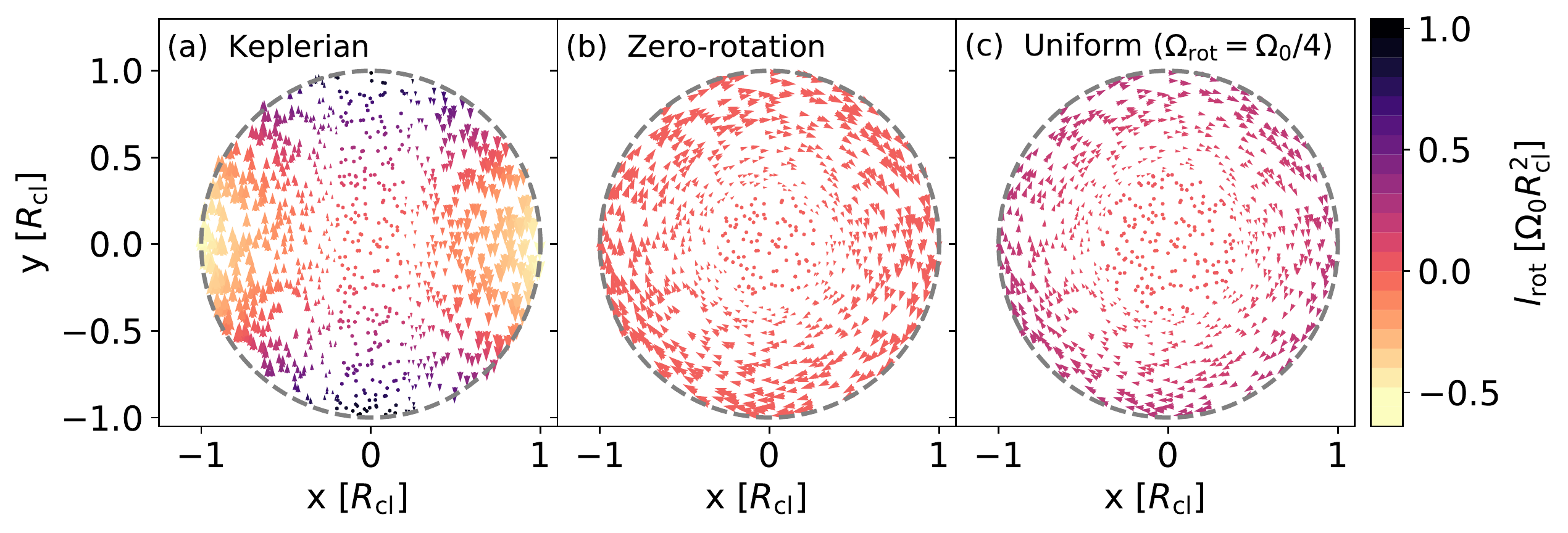}
\caption{Quiver plot of initial velocities in the co-moving frame for three setups used in this paper. While our simulations use spherical clouds, this plot shows their cross-sections at z=0. The arrow sizes correspond to the magnitude of the particle velocities and the colors correspond to the rotational contribution (from Eq. \ref{eq:transformation}). The left panel (a) indicates the Keplerian setup (\ref{eq:Keplerian}), which yields a total prograde rotation equal to $L_\mathrm{rot} \simeq 0.1 M_\mathrm{cl} \Omega_0 R_\mathrm{cl}^2$. The middle panel (b) corresponds to the zero-rotation initial condition (Eq. \ref{eq:zero-rotation}), while the right panel (c) corresponds to uniform rotation with the same initial rotational angular momentum as the Keplerian setup.}
\label{fig:arrowplots}
\end{figure*}

In local simulations such as the co-moving frame with/without shearing sheet approximation, particle positions and velocities are determined in a frame that is both rotating and translated relative to the star. In order to calculate the rotation vector of a cloud from these quantities, it is necessary to transform the vectors correctly to an inertial frame with the origin located on the studied object. The results of such a transformation for circular orbits are given in 3D without derivation by \citet{Nesvorny2019} and in 1D by \citet{Giuli1968, DonesTremaine1993}. As far as we are aware, only a limited 1D justification is shown by \citet{LissauerKari1991}. Because we could not find a complete 3D derivation of the required transformation anywhere, we provide it here.

Let the first, co-moving frame with Cartesian coordinates ($x,y,z$) be centered at a (vector) distance $\mathbf{r_0}$ of a central object, in a frame that is rotating around the central object with the rotation vector $\mathbf{\Omega_0} = \Omega_0 \hat{z}$. Let the second frame with Cartesian coordinates ($x',y',z'$) share its origin with the first frame but be stationary (inertial). In order to transform a vector from the first to the second frame, the following operations are required:
\begin{enumerate}
    \item Translate the frame, such that the origin coincides with the center of the orbit: $\mathbf{r} \rightarrow \mathbf{r + \mathbf{r_0}}$.
    \item Rotate the frame around the origin with $-\Omega_0$, such that the frame becomes inertial: $\mathbf{v} \rightarrow  \mathbf{v} + \mathbf{\Omega_0 \times r}$.
    \item Translate the frame back, such that the origin is again located at the center of the cloud: $\mathbf{r + \mathbf{r_0}} \rightarrow \mathbf{r}$.
\end{enumerate}
Together, these transformations lead to the following distance and velocity vectors in the stationary frame:
\begin{subequations}
\begin{align}
    \mathbf{r'} &= \mathbf{r} \\
    \mathbf{v'} &= \mathbf{v} + \mathbf{\Omega_0 \times (r + \mathbf{r_0})}. \label{eq:v_transform}
\end{align}
\end{subequations}
The rotation of a particle is given by its velocity relative to the cloud's center-of-mass. If the center-of-mass remains located on $\mathbf{r_0}$, it corresponds to 
$\mathbf{v'_\mathrm{cm} = \Omega_0 \times \mathbf{r_0}}$ and the specific rotational angular momentum of a particle is given by:
\begin{subequations}
\begin{align}
    \mathbf{l'_\mathrm{rot}} &= \mathbf{r' \times (v' - v'_\mathrm{cm}) } \\
    &= \mathbf{r \times (v + \Omega_0 \times r)} \label{eq:subeq_omega} \\
    &= \begin{pmatrix}
        yv_\mathrm{x}-zv_\mathrm{y}  - \Omega_0 xz\\\
        zv_\mathrm{x}-xv_\mathrm{z}  - \Omega_0 yz \\
        xv_\mathrm{y}-yv_\mathrm{x}  + \Omega_0 (x^2+y^2) \end{pmatrix}. \label{eq:transformation}
\end{align}
\end{subequations}
This final equation can trivially be generalized to any direction of $\mathbf{\Omega_0}$ by using non-zero xy-components in Eq. \ref{eq:subeq_omega}. We note that there is, however, a potential issue with this calculation in a co-moving frame. If the center-of-mass of the cloud shifts inward by a significant distance, the transformation no longer uses the right reference point for the rotation, as the object forms on a slightly interior orbit. The magnitude of this error grows with $\delta R/r_0 \propto \left(R_\mathrm{cl}\Omega_0 t/r_0\right)^2$ (see Eq. \ref{eq:COM_shift}) and is not quantitatively important in the calculations performed in this work.

%%%%%%
\section{Initial rotational condition of spherical clouds}\label{Appendix:rotation_IC}
In Sects. \ref{sect:NumCC} and \ref{sect:ApplCC}, we perform collapse simulations with spherical clouds that start with three different velocity setups, which we explain here. We refer to the first setup we use as \textit{Keplerian}, meaning that all particles in the cloud begin with velocities in the stationary frame (whose axes directions at $t=0$ coincide with a co-rotating frame) equal to:
\begin{equation}\label{eq:Keplerian}
    \mathbf{v'_\mathrm{Keplerian}}(t=0) = \mathbf{\Omega} \times \mathbf{r}.
\end{equation}
We show these velocities translated to a co-moving frame in panel (a) of Fig. \ref{fig:arrowplots}. In this frame, the velocity differences are mainly visible as shear ($v_\mathrm{y} \simeq -\frac{3}{2} \Omega_0 x$) on the x-axis. From Eq. \ref{eq:transformation}, it is easy to see that the rotational contribution of particles on the x-axis becomes retrograde due to this shear, ranging from 0 to $-0.5 \Omega_0 R_\mathrm{cl}^2$. On the y-axis, however, the shear is negligible and absolute velocity differences are minimal. Nevertheless, the curvature of the orbit (captured in the term $\Omega_0 (x^2+y^2)$) provides significant directional differences in velocity, which lead to prograde rotational contributions ranging from 0 to $\Omega_0 R_\mathrm{cl}^2$. When integrated over the complete sphere, the net initial rotational condition of the Keplerian setup is around $L_\mathrm{rot} \simeq 0.1 M_\mathrm{cl} \Omega_0 R_\mathrm{cl}^2$.

We refer to the second setup used in this work as the 
\textit{zero-rotation} condition. In this setup, we set all the velocities equal to
\begin{equation}\label{eq:zero-rotation}
    \mathbf{v'}_\mathrm{zero-rotation}(t=0) = \mathbf{\Omega_0} \times \mathbf{r_0},
\end{equation}
such that all particles begin with the same velocity as a particle that is located at the sphere's center-of-mass and orbits in a circular motion. Without velocity differences, the particles in the cloud have no initial rotation. However, this is not immediately visibly apparent in the co-rotating frame (panel (b) of Fig. \ref{fig:arrowplots}), where the velocities are transformed (Eq. \ref{eq:v_transform}). Even though the particles provide no physical rotational angular momentum, their rotation visibly appears to start in a retrograde condition in this frame with a rotational period equal to the orbital period. 

Finally, the third setup we use in this paper is one with a uniform cloud rotation rate ($\omega_\mathrm{rot}$) where the initial velocities are set equal to:
\begin{equation}\label{eq:uniform}
    \mathbf{v'}_\mathrm{uniform}(t=0) = \mathbf{\Omega_0} \times \mathbf{r_0} + \mathbf{\omega_\mathrm{rot}} \times (\mathbf{r}-\mathbf{r_0}).
\end{equation}
The initial rotational condition in this setup trivially follows as $L_\mathrm{rot} = I_\mathrm{cl} \omega_\mathrm{rot} = 0.4 M_\mathrm{cl} \omega_\mathrm{rot} R_\mathrm{cl}^2$. In Fig. \ref{fig:arrowplots}, we plot the velocities in a co-rotating frame for the case where $\omega_\mathrm{rot} = \Omega_0/4$ and the initial rotational angular momentum is identical to the Keplerian setup. In the special case where $\omega_\mathrm{rot}=\Omega_0$ is chosen, all the particles appear stationary in the co-moving frame.

\section{Collapse of a spherical cloud in a co-moving frame}\label{appendix:collapse_panels_corot}
\begin{figure*}[t] 
\centering
\includegraphics[width=\hsize]{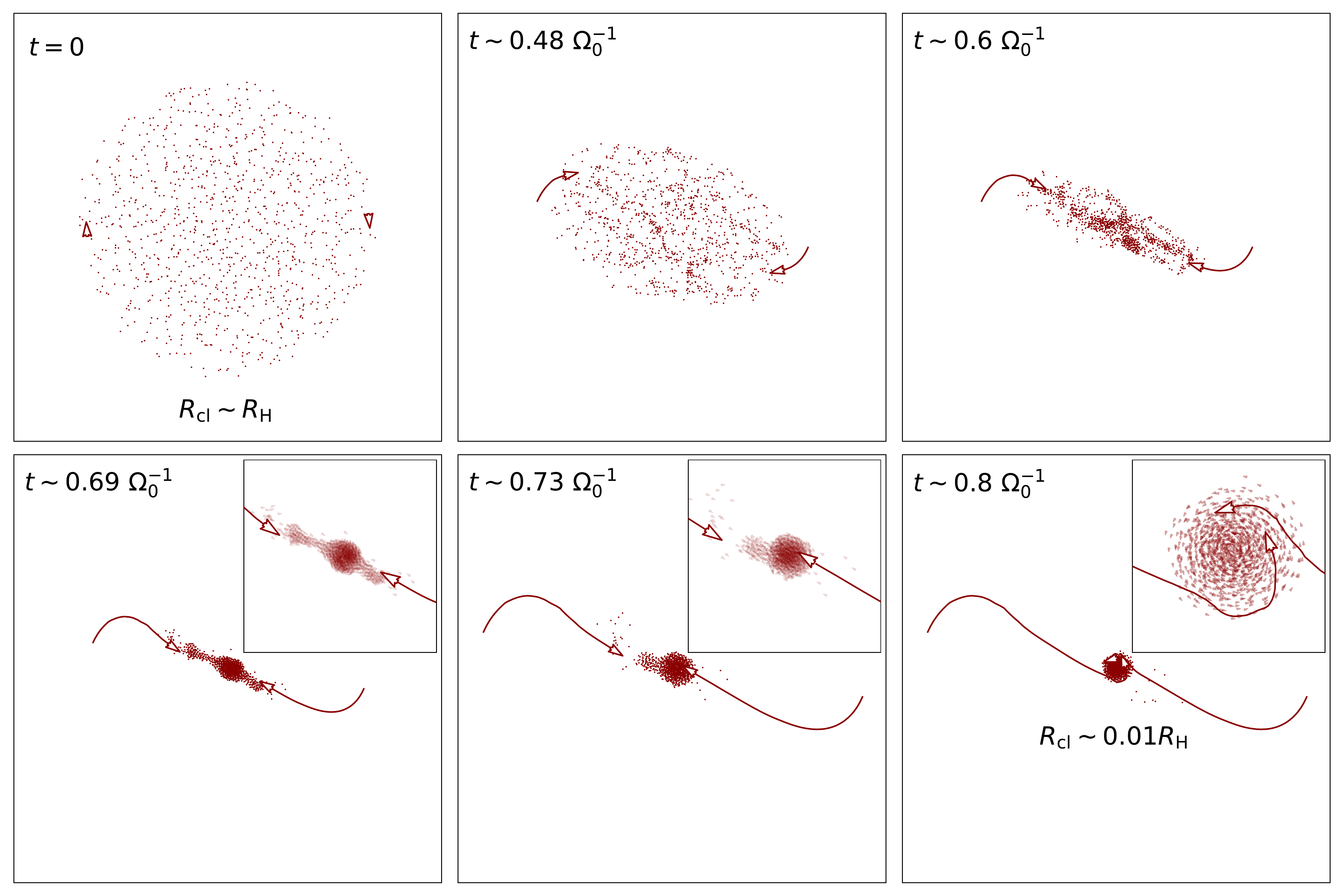}
\caption{Overview of the collapse of a uniform cloud in orbit around a central mass (same as Fig. \ref{fig:collpanelshelio}), shown in a co-moving frame. The spherical cloud is initiated with size $R_\mathrm{H}$ and without any rotation. The central mass is located left from the cloud. The velocity directions (magnitudes not to scale) and trajectory of the particles are highlighted for an interior and exterior particle, that wrap around the center-of-mass in a prograde fashion.}
\label{fig:collpanels}
\end{figure*}
In Fig. \ref{fig:collpanels}, we repeat the collapse of an initially non-rotating spherical cloud in orbit (same calculation as Fig. \ref{fig:collpanelshelio}), visualized instead in the commonly used co-rotating frame - rather than a stationary one. In this frame, the curvature of the orbital motion that is naturally present in a stationary frame is instead introduced by the Coriolis acceleration. In the first panel, the non-rotating cloud visually appears to have a retrograde rotation, which is a consequence of the non-inertiality of the frame (see also Fig. \ref{fig:arrowplots}(b)). As the cloud shears out and contracts, it again first attains a bar-like shape. When the particles are pulled toward the center-of-mass, they wrap around it in a prograde fashion. Finally, the collapse finishes and a spherical object with prograde rotation has formed. The physical shape and multiplicity of the system that forms, depends on the numerical parameters assumed, such as the filling factor and number of numerical particles. Importantly, however, the rotational direction and its magnitude remain unaffected by these numerical parameters.
}
\end{document}